\documentclass[runningheads]{llncs}
\usepackage{array,multirow}
\usepackage{graphicx}
\usepackage{subfig}
\usepackage{xcolor}
\usepackage{cite}
\usepackage{mathtext}
\usepackage{amsmath}
\usepackage{amssymb}

\DeclareMathOperator{\argmin}{arg\,min}
\renewcommand{\v}[1]{{\mathbf{#1}}}

\begin{document}

\title{Optimizing Unlicensed Coexistence Network Performance Through Data Learning}
\titlerunning{Coexistence Network Optimization Through Data Learning}
%
\author{Srikant Manas Kala\inst{1}
 \and
Vanlin Sathya\inst{2}
 \and
Kunal Dahiya\inst{3}
\and
Teruo Higashino\inst{1}
\and
Hirozumi Yamaguchi\inst{1}
}
\authorrunning{Srikant Manas Kala et al.}
%
\institute{Mobile Computing Laboratory, Osaka University, Japan\\
\email{\{manas\_kala, higashino, h-yamagu\}@ist.osaka-u.ac.jp}
\and
The University of Chicago, Illinois, USA\\
\email{vanlin@chicago.edu}\\
\and
Indian Institute of Technology Delhi, India\\
\email{kunalsdahiya@gmail.com}}

\maketitle              
\begin{abstract}
Unlicensed LTE-WiFi coexistence networks are undergoing consistent densification to meet the rising mobile data demands. With the increase in coexistence network complexity, 
it is important to study network feature relationships (NFRs) and utilize them to optimize dense coexistence network performance.
This work studies NFRs in unlicensed LTE-WiFi (LTE-U and LTE-LAA) networks through supervised learning of network data collected from real-world experiments. Different 802.11 standards and varying channel bandwidths are considered in the experiments and the learning model selection policy is precisely outlined.
 Thereafter, a comparative analysis of different LTE-WiFi network configurations is performed through learning model parameters such as R-sq, residual error, outliers, choice of predictor, \emph{etc.} Further, a Network Feature Relationship based Optimization (NeFRO) framework is proposed. NeFRO improves upon the conventional optimization formulations by utilizing the feature-relationship equations learned from network data. It is demonstrated to be highly suitable for time-critical dense coexistence networks through two optimization objectives, \emph{viz.,} network capacity and signal strength. NeFRO is validated against four recent works on network optimization. NeFRO is successfully able to reduce optimization convergence time by as much as 24\% while maintaining accuracy as high as 97.16\%, on average.

\keywords{LTE-WiFi Coexistence  \and Network Optimization \and Machine Learning.}
\end{abstract}
\section{Introduction}
 Cellular networks are a vital component of a truly mobile augmented reality (AR) system/application such as ``Pokemon Go," as they offer the widest coverage to the end-users. With the rising demand for immersive AR experience, the AR market is set to cross \$100 Billion and the total mobile network traffic is  expected to exceed 300 Exabytes per month in 2026 \cite{EriccsonMobilityReport2021}.
 
 However, mobile AR traffic is latency-critical, uplink-heavy, and bursty in nature and the current LTE/LTE-A (Long Term Evolution/Long Term Evolution-Advanced) networks lack the capability to offer a seamless mobile AR experience \cite{ARLTE}.
  Consequently, cellular operators have taken several measures such as dense deployment of small-cells (SCs) and access points (APs) and utilization of the unlicensed spectrum through LTE-WiFi coexistence.

 The prospect of effectively utilizing the unlicensed spectrum through \textit{LTE in unlicensed spectrum} (LTE-U) and \textit{LTE license assisted access} (LTE-LAA) appeals to the mobile operators. Hence, there is a rapid deployment of both LTE small-cells and Wi-Fi APs in the 5GHz band where 500 MHz of the unlicensed spectrum is shared by both LTE and Wi-Fi networks \cite{ACM,icdcn}. 
 
 This work focuses on two aspects of LTE-WiFi coexistence 
\emph{viz.,} coexistence network performance analysis and time-critical optimization. To that end, a comparative performance analysis of unlicensed LTE standards (LTE-U/LAA) is done through network feature relationship parameters learned from network data. Thereafter, the learned feature relationships are utilized to reduce the time-cost of performance optimization in a dense coexistence network.

\subsection{Motivation}
With the proliferation of unlicensed coexistence networks, there has been a significant debate on the comparison of LTE-U and LAA standards and their performance. While cellular operators such as AT\&T and Verizon have opted in favor of LAA deployments \cite{ACM}, recent works claim that LTE-U may offer better coexistence with Wi-Fi under specific conditions \cite{UvsLaa}. 


The existing comparative studies of LTE unlicensed standards are lacking in three respects. First, they primarily rely on simulations and make several assumptions \cite{LTEvsWiFi, UvsLaa}. Secondly, the offered comparative analysis is based only on \textit{measurements}, \emph{i.e.,} by simply comparing several network performance evaluation metrics such as throughput, latency, number of re-transmissions, \emph{etc.} In contrast, \textit{feature relationship analysis} looks for patterns in network data that can reveal relationships between network variables such as dependence, correlation, causation, \emph{etc.}  Finally, the variation in performance of LTE unlicensed variant with the variation in coexisting Wi-Fi standard is often overlooked. In addition, the impact of factors such as bandwidth allocation and signaling data is rarely studied. 


With the increase in the deployment of small-cells and access points, dense networks (DNs) with inter-site distance $\leq$ 10m, and ultra-dense networks (UDNs) with inter-site distance $\leq$ 5m, have proliferated in most urban centers \cite{dense2}. Thus, performance optimization of the rapidly growing dense coexistence networks is a major challenge. This becomes particularly important when time-critical mobile AR services/applications need to be supported by coexistence networks. However, the literature currently lacks network feature relationship (NFR) analysis from the perspective of dense LTE-WiFi coexistence networks. Further, to the best of our knowledge no existing study makes use of network feature relationships in dense coexistence network optimization. 

 \subsection{Contributions}
 In this work, we address these concerns through the following contributions
 \begin{itemize}
     \item Study network feature relationship in dense coexistence networks such as SINR-Capacity relationship, through machine learning algorithms.
     \item Analyze the impact of factors such as the choice of LTE unlicensed standard, coexisting Wi-Fi standard, and bandwidth allocation 
     on NFRs in coexistence networks.
     \item Compare LTE-LAA/LTE-U and Wi-Fi 802.11n/ac coexistence performance based on NFR parameters such as the choice of predictor variable, R-sq (model validity), residual error (absolute and normalized), outliers, \emph{etc.}
     \item Utilize NFRs to optimize dense coexistence network performance through network capacity and signal strength optimization.
 \end{itemize}

 The comparative analysis in this work is distinct from the state-of-the-art studies \cite{LTEvsWiFi, UvsLaa} in that it is not limited to measuring and analyzing individual network variables. It involves data-learning to discover feature relationship patterns which determine network performance. Further, the data is gathered through real-time experiments instead of simulations. For the experiments, dense and ultra-dense co-existence networks were implemented with the help of USRP NI-SDRs and WiFi APs. 
 The \textit{learning model selection policy} considered for feature relationship analysis is also explicitly described for replication and validation. 
  \begin{figure}[ht]
                \centering
                \includegraphics[width=0.9\linewidth,trim=5mm -5mm 0mm -3mm]{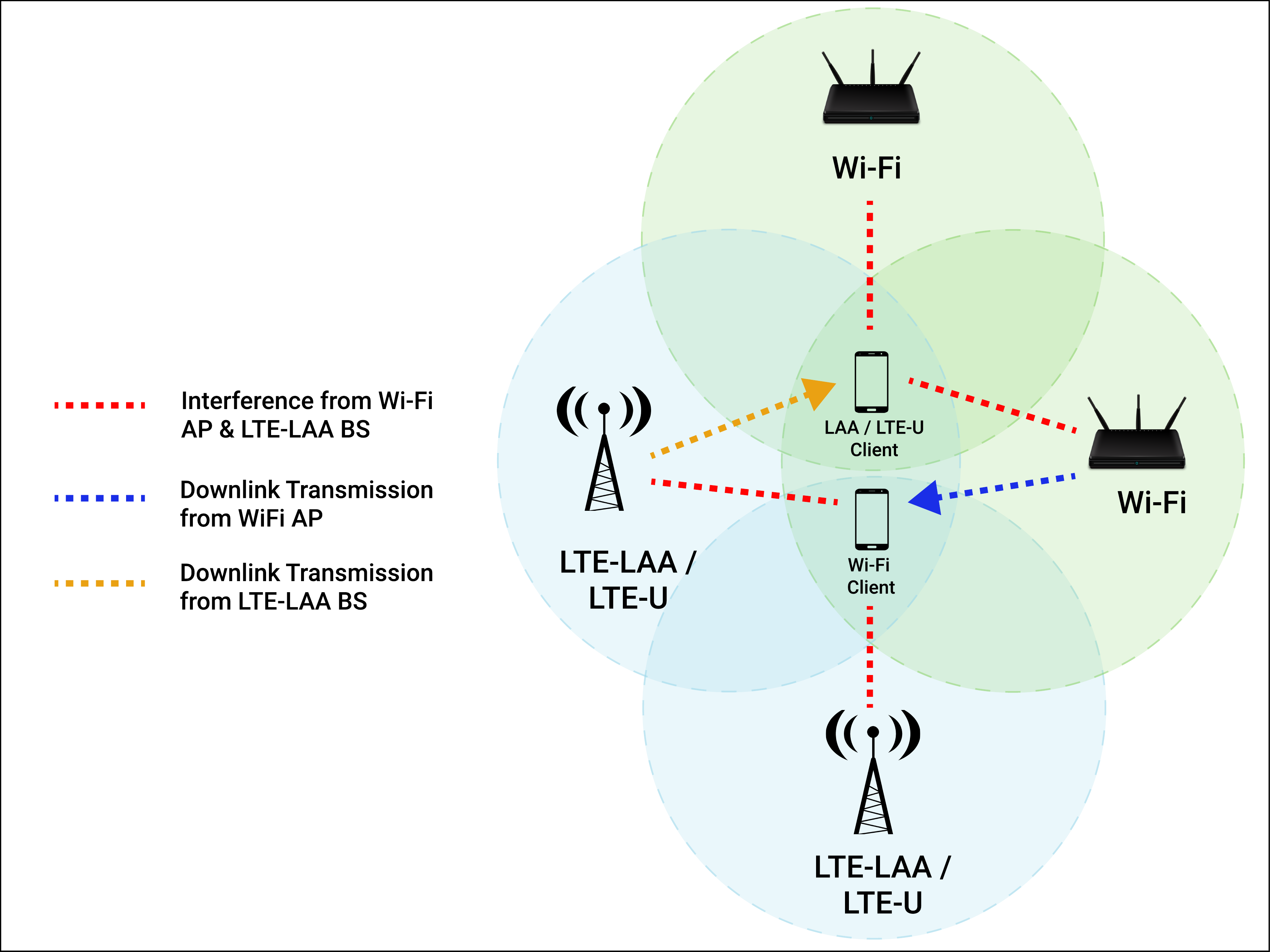}
                \caption{Interference in Dense Unlicensed Coexistence Networks}
                \label{LTEINT}
        \end{figure}
 \section {A Review of Related Works} \label{related}
 
\subsection{Network Feature Relationships in Dense Networks}
 In the recent past, several state-of-the-art studies have used regression algorithms, decision trees, and other machine learning techniques for NFR analysis \cite{regmswim, regsigcom, regwcnc, icdcn}.
Some of these works leverage the learned NFRs to improve network performance. For example, learning 802.11n feature relationships can facilitate improved configuration selection and enhanced rate adaption \cite{regmswim}. Yet, the current literature lacks a robust analysis of NFRs, such as the capacity-interference relationship (CIR) in unlicensed coexistence networks.

Further, as shown in Figure~\ref{LTEINT}, densification of LTE-WiFi coexistence systems will exacerbate the adverse impact of interference and pose additional challenges. While densification may lead to an initial gain in LTE-WiFi coexistence system capacity, network performance eventually deteriorates with rise in density \cite{CoexDense1}.   
  Moreover, the impact of factors \emph{e.g.,} unlicensed LTE variant, Wi-Fi standard, bandwidth allocated, and signaling data, \emph{etc.,} on dense coexistence CIR also remains unexplored. For example, the analysis presented in \cite{smkcomsnets} is limited to demonstrating how the SINR-Capacity relationship differs in regular and dense/ultra-dense networks, and fails to explore the impact of the factors listed above.
 
  Therefore, this work focuses on various aspects of the relationship between interference and network performance in a dense coexistence network.
\subsection{Optimization Challenges in Dense Networks}

The need for low association times and fast-handovers in a dense environment makes network optimization \textbf{time-critical}. However, consistent densification significantly increases network scale and complexity which leads to high convergence times and computational overhead to arrive at optimal solutions \cite{dense2}. 
This is a major challenge for ultra-low-latency AR applications as already the LTE/LTE-A deployments account for almost 30\% of the end-to-end AR latency \cite{ARLTE}. With densification, the latency problem will exacerbate and diminish the gains in throughput. 

 Thus, it is important not only to study the impact of densification on NFRs but also ascertain how these feature relationships can be used to accelerate optimization in dense coexistence networks by making it computationally less expensive \cite{OptCIR}.
Broadly speaking, wireless network performance can be optimized through three major frameworks \emph{viz.,} optimization, machine learning, and a hybrid approach that involves machine learning based optimization \cite{MLOPT, OptCIR}. 

This work paves the way for an empirical and practical approach to \textit{\textbf{network feature relationship based optimization}} (NeFRO). NeFRO adopts the hybrid model wherein feature relationships learned from network data serve as a constraint in network optimization formulations. 
By using the feature relationship equation for performance optimization, NeFRO accounts for the ambient network environment and is free from theoretical pre-suppositions. Due to these factors, NeFRO is shown to significantly reduce the time-costs in dense network performance optimization.

\section{Experimental Set-up}
This section describes the experimental platform designed to create a dense LTE-WiFi coexistence environment in the 5GHz unlicensed spectrum. The testbed is used to collect data for NFR analysis.
\begin{figure}[h!]
                \centering
                \includegraphics[width=0.95\linewidth,trim=5mm -5mm 0mm -3mm]{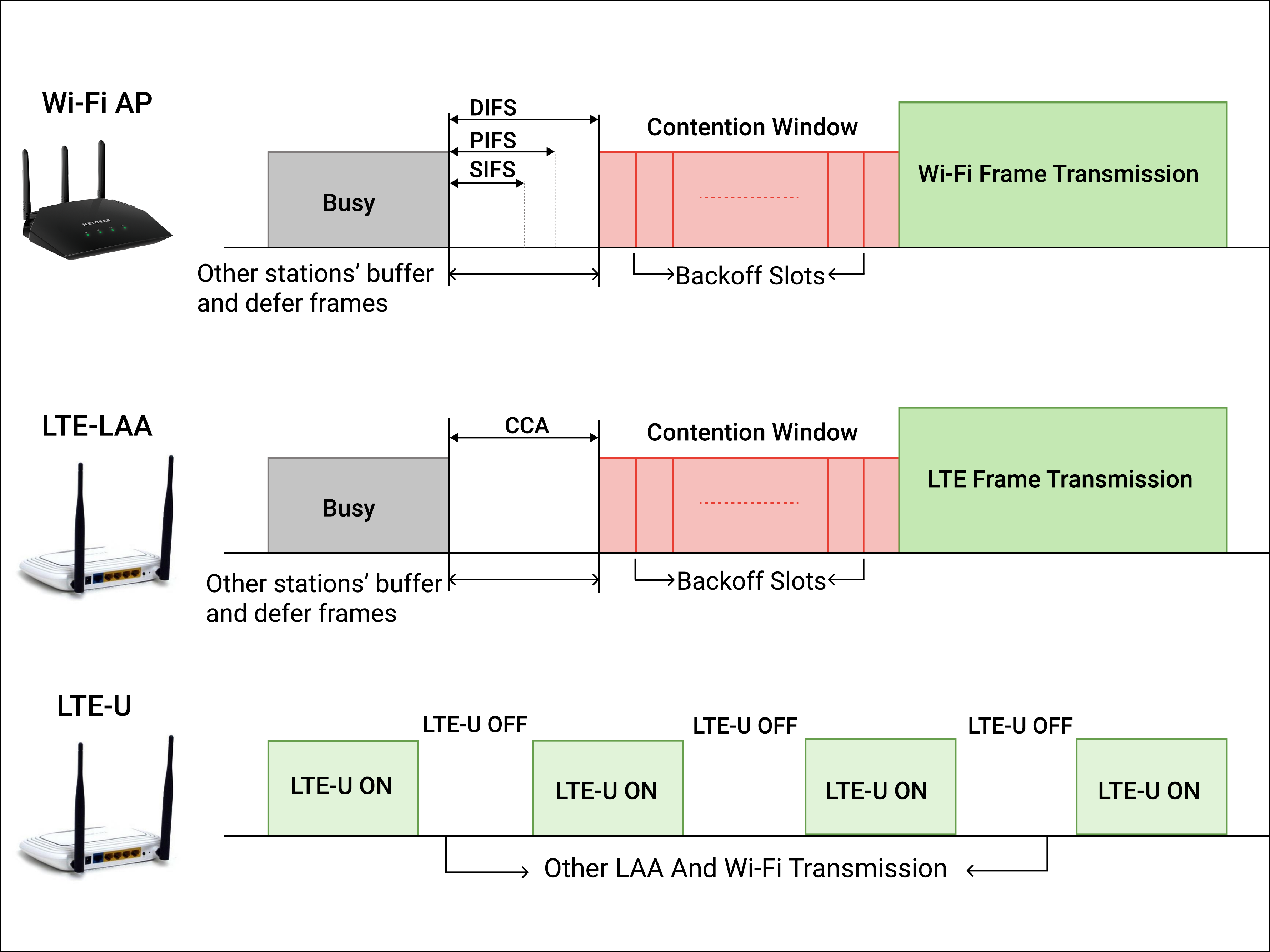}
                \caption{Wi-Fi, LTE-LAA, and LTE-U: Channel Access Mechanisms}
                \label{COMAC}
        \end{figure}
\subsection{Testbed Design}
Two variants of LTE unlicensed operation have been standardized and released, viz., LTE-U and LTE-LAA, albeit with starkly different medium sensing and access mechanisms. LTE-U relies on a load-dependent duty-cycle mechanism based on Carrier Sense Adaptive Transmission (CSAT). On the other hand, LTE-LAA depends on a Listen-Before-Talk (LBT) mechanism which is similar to the CSMA/CD MAC protocol of Wi-Fi, making it relatively easier for LAA to coexist with IEEE 802.11 WLANs. The medium access mechanisms of the two LTE unlicensed variants and Wi-Fi are juxtaposed in Figure~\ref{COMAC}.

\par\textbf{LAA-LTE/LTE-U Platform} The National Instruments \textit{NI RIO} testing-platform is used as the LAA/LTE-U testbed as shown in Figure~\ref{exp}~(a). The PHY on the NI Labview system is the standard PHY implementation as prescribed in the LTE-A 3GPP release. More technical details on the testbed are presented in Table~\ref{sim}. The system offers high operational flexibility through advanced user-defined configuration of signal transmission and reception. Several network parameters can be configured, such as the sub-carrier modulation scheme, resource block allocation, LAA transmission opportunity (TXOP), Energy Detection (ED) threshold, LBT category option, LTE-U duty cylce ON \& OFF, transmission power, OFDM parameters (\emph{e.g.,} 1 to 3 control channels), carrier frequency offset, and timing offset estimation.
\par\textbf{Wi-Fi Platform} Netgear wireless routers
are used to design the Wi-Fi testbed. The off-the-shelf
Wi-Fi routers, supporting both 802.11n and 802.11ac in the 5 GHz band serve as the typical Wi-Fi nodes. The Wi-Fi testbed supports easy modification and monitoring of parameters and functions in both the MAC and PHY layers of Wi-Fi such as DIFS, CWmin, CWmax, channel bandwidth, and transmission power.

\subsection{Experiment Design}
All experiments are carried out in the typical setting of an indoor office at the University of Chicago campus. This work focuses mainly on gathering SNR and throughput data for NFR analysis. Other network parameters such as contention window size, request to send (RTS), clear to send (CTS), inter-beacon interval time, power range, channel assignment (static or dynamic), and bandwidth in the PHY layer are also configured as required. In the experiments, the LAA transmitter always uses LBT protocol to sense if the channel is available and the maximum TXOP is 8 ms, which is similar to the transmission of LTE-A in licensed bands. The Power Spectral Density (PSD) of LAA transmissions is controlled so as to ensure that the power of the interference from LAA is below Clear Channel Assessment (CCA) threshold of Wi-Fi communications.
\begin{figure*}[ht!]
 \centering%
\begin{tabular}{cc}
	\subfloat[Representative Testbed] {\includegraphics[width=.5\linewidth]{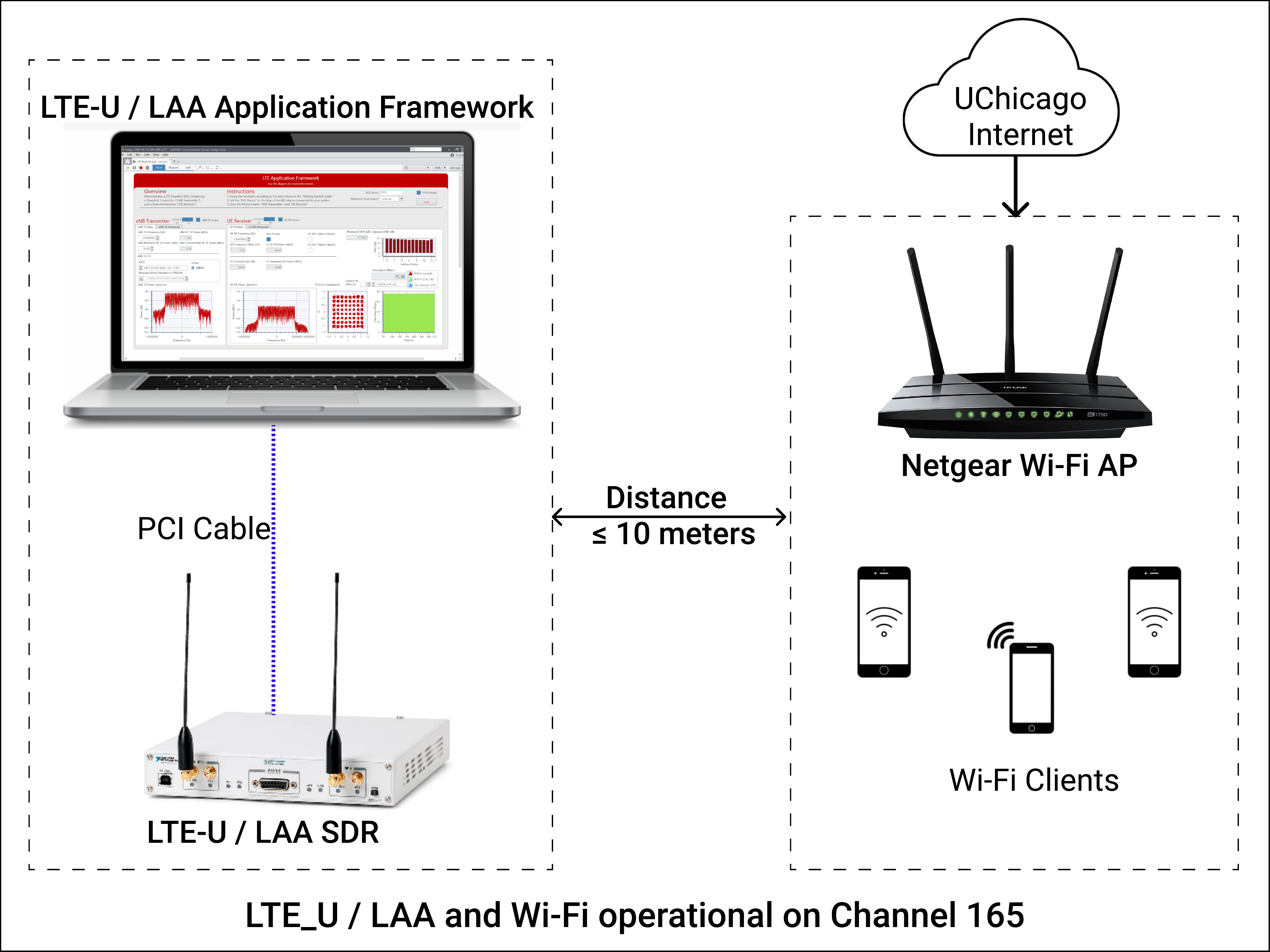}}\hspace*{0.1cm}\hfill%
    \subfloat[Representative Topology] {\includegraphics[width=.40\linewidth]{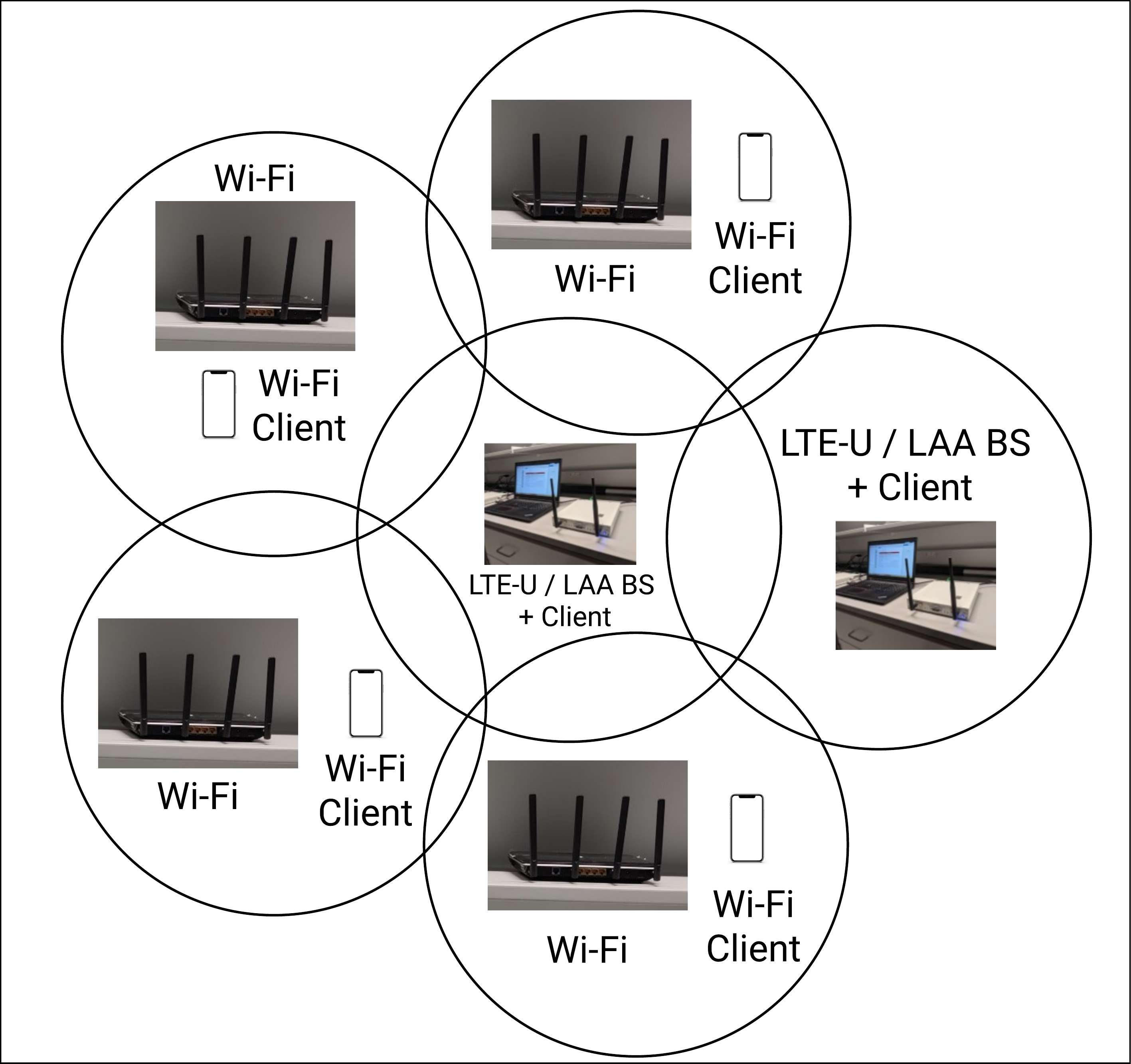}}\hspace*{0.1cm}\hfill%
\end{tabular}
  \caption{Experimental Set-up} 
    \label{exp}
\end{figure*}
Several experiments were designed to explore dense unlicensed coexistence performance by creating combinations of LAA/LTE-U, 802.11n/802.11ac, and different bandwidths (5/10/15/20 MHz). LAA and LTE-U use the same underlying mechanism of Dynamic Bandwidth Adaptation (DBA) for spectral efficiency as LTE-A. Therefore, while Wi-Fi APs generally operate in a bandwidth of 20 MHz, LAA and LTE-U possess the capability to support multiple bandwidths (1.4/3/5/10/15/20 MHz). Bandwidth is an important factor that may influence capacity interference relationship due to cross-talk interference. Therefore, this work considers bandwidth to be an important parameter for CIR analysis. 
Further, dense random topologies are considered where LAA/LTE-U/Wi-Fi nodes are placed at inter-nodal distances of 5m to 10m. A representative illustration is presented in Figure~\ref{exp}~(b). Apart from a small inter-nodal distance, a dense coexistence scenario in an indoor setting is also interesting due to the prevalence of significant multi-path fading and presence of obstacles such as walls, furniture, objects, etc. 

\begin{table}[h]\centering
\caption{Experiment Parameters}
\resizebox{0.65\textwidth}{!}{\begin{tabular}{|c|c|}
\hline
\textbf{Parameter} 	   & \textbf{Value} 
	 \\ \hline
Number of nodes 
 & 6  \\ \hline 
Transmission Power     & 23 dBm \\ \hline
Operating Frequency        & 5 GHz \\ \hline
LTE-U/LAA RF Transmission        & Loopback \\ \hline
LTE Transmission Channel & PDSCH, PDCCH \\ \hline
Data Traffic    & Full buffer \\ \hline
Wi-Fi Channel Access Protocol   & CSMA \\ \hline
LAA Channel Access Protocol   & LBT \\ \hline
\multicolumn{2}{l}{\footnotesize *PDSCH - Physical Downlink Shared Channel} 
\end{tabular}}
\label{sim}
\end{table}

\section{Network Feature Relationship Analysis Methodology}

Regression is a popular machine learning paradigm used to determine the relationship between network parameters in continuous space~\cite{regmswim, regsigcom, regwcnc, icdcn}.
Regression algorithms not only offer reliable feature relationships, but also provide insights into the relationship in terms of model validity, outliers, residual error \emph{etc.} Thus CIR is modeled as a bi-directional regression problem where the goal is to estimate or predict network capacity through SINR feature points, and vice versa.

\subsection{Learning Algorithms for Relationship Analysis}
 Let $N$ represent the number of training points and let dimensionality of the feature vector be denoted by $D$. Then, the coexistence network data can be represented as $\{\v x_i, y_i\}_{i=1}^N$, where $\v x_i \in \mathbb{R}^D$ is the feature vector and $y_i \in \mathbb{R}$ is the ground truth value for $i^{th}$ training point. The goal is to learn a mapping $f: \v x_i \xrightarrow{} y_i$ where $x_i$ is the predictor (SINR or Capacity) and $y_i$ is the response (Capacity or SINR).
This work considers the following basket of learning algorithms for the regression analysis:
\begin{itemize}
\item \textbf{Linear Regression} This group of algorithms learns a linear relationship by solving $\argmin_{\v w, b} \sum_{i=1}^N ||(\v w^{\top} \v x_i+b) - y_i||_2^2 + \alpha\v w^{\top}\v w$~\cite{murphy2012machine}. Here, the weight vector is denoted by $\v w \in \mathbb{R}^D$ and the bias term is $b \in \mathbb{R}$. Further, the weightage (importance) of the $l_2$-regularization term is controlled by the hyper-parameter denoted by $\alpha$, which is set to zero for Ordinary Least Squares Linear Regression~(OLS). However, for Ridge Regression~(RR), $\alpha$ is set through $k$-fold cross validation (kCV).

\item\textbf{Kernel Ridge Regression} A non-linear mapping is expected to be more suitable for the SINR-Capacity relationship \cite{Manas}. Therefore, we make use of the Kernel Ridge Regression~\cite{murphy2012machine} that employs non-linear transformations such as Polynomial and Radial Basis Function~(RBF). Its goal is to solve $\argmin_{\v w, b} \sum_{i=1}^N ||K(\v w, \v x_i)+b - y_i||_2^2 + \alpha\v w^{\top}\v w$. Here, $\v w \in \mathbb{R}^D$ is the weight vector, $b \in \mathbb{R}$ is the bias term, and $\alpha$ is a hyper-parameter defined above. Finally, $K(a, b)$ is a kernel function which allows to compute dot product in an arbitrary large space without the need to explicitly project features in high dimensional space. 
Varying the kernel function as RBF and Polynomial leads to Kernel RBF Regression~(RBF) and Multi-variate Polynomial Regression~(MPR), respectively.
\end{itemize}

\subsection{Selection of Regression Models}

Regression Model selection depends upon objective criteria such as R-sq, higher-order terms, \emph{etc.,} and some subjective value-judgments, \emph{e.g.,} selecting a model with a higher R-sq even if the higher-order terms are not significant. However, studies often discuss network feature relationships and existence of correlation without going into the details of the underlying regression models \cite{regsigcom}. Failure to highlight such details poses a challenge while replicating these studies. To avoid this problem, the model selection policy considered in this work is described below.

\par\textbf{Regression Model Selection Policy} \label{RMSP}
The regression algorithms are subjected to \textit{k-Fold Cross-validation (kCV)} averaged over 30 runs (for $k=5$). Feature relationship models are evaluated based on their R-sq or \textit{Regression Model Validity} (RMV). A high RMV value signifies the \textit{goodness} of the fit. Also, outlier detection and removal is performed using the Local Outlier Factor (LOF) algorithm.

First, CIR models with 1--3 degree polynomials are learned and to avoid over-fitting of feature point data, the higher-order terms considered are limited to statistically significant cubic terms. Further, the optimal model is chosen on the basis of RMV via kCV as it best explains the feature relationship \cite{Reg2}. For example, between a CIR model learned from the baseline data-set and the model learned from the data-set processed through LOF outlier removal, the model and feature relationship with the higher RMV is considered.
 This work focuses primarily on quadratic CIR models for the following reasons. First, the \% difference in average RMVs of linear \& quadratic and quadratic \& cubic models is 3.63\% and 0.98\% respectively. Thus, as compared to quadratic models, the linear models exhibit a relatively weak CIR and the RMV gain in cubic models is very low. Second, CIR in wireless networks is expected to be quadratic \cite{2Gupta}. Finally, low convergence time is a primary constraint in dense network optimization. Whatever little gain the higher RMV of a cubic model might offer in performance optimization, will be irrelevant compared to the increase in the computational overhead of a third-degree polynomial constraint. 

\subsection{Analytical Methodology}
To study the impact of dense network configuration on NFRs, it is necessary to isolate individual network parameters and observe the consequent variation in the feature relationship. 
\par\textbf{Comparative Themes}
The analysis seeks to draw a comparison between the performance of LTE unlicensed variants (LTE-U and LTE-LAA) in coexistence with the Wi-Fi variants (802.11n/ac). We also study the impact of bandwidth allocation and the choice of predictor variable on CIR in these network configurations. 
Thus, a total of 32 Test Scenarios are considered (denoted by TS$_{i}$, where $i\in \{1\ldots32\}$). Each TS$_{i}$ indicates a unique unlicensed coexistence network scenario based on the LTE unlicensed variant (LTE-U/LTE-LAA), coexisting Wi-Fi standard (802.11n/ac), bandwidth allocated (5/10/15/20 MHz), and predictor variable (SINR/Capacity). For each TS$_{i}$, the CIR model is selected through the regression model selection policy outlined earlier.


\par\textbf{Comparison Parameters}
The performance of different LTE-WiFi network configurations is evaluated through analysis of learning parameters such as model validity, standard deviation in RMV, residual standard deviation (RSD), outliers, \emph{etc.}. Trends of average network values observed in the experiments are used as well. 
For each of these parameters, two types of comparisons are carried out, \emph{viz.} scenario-specific comparison and component-specific comparison for LTE-WiFi-Predictor configurations. The former is aimed at a comparative analysis of individual network scenarios (\emph{e.g.,} LTE-U, 802.11n, at 5MHz vs. LTE-LAA, 802.11n, at 5MHz ) while the second is aimed at capturing component level trends (\emph{e.g.,} SINR as a predictor vs. Capacity as a predictor). Reliable inferences are drawn only if the findings are consistent at both levels of comparative analysis. Wherever possible, plausible explanations are offered.

\section{CIR in Dense Unlicensed Coexistence Networks} \label{CIRDCN}

\begin{figure*}[ht!]
 \centering%
\begin{tabular}{cc}
    \subfloat[LTE-LAA vs. LTE-U] {\includegraphics[width=.45\linewidth]{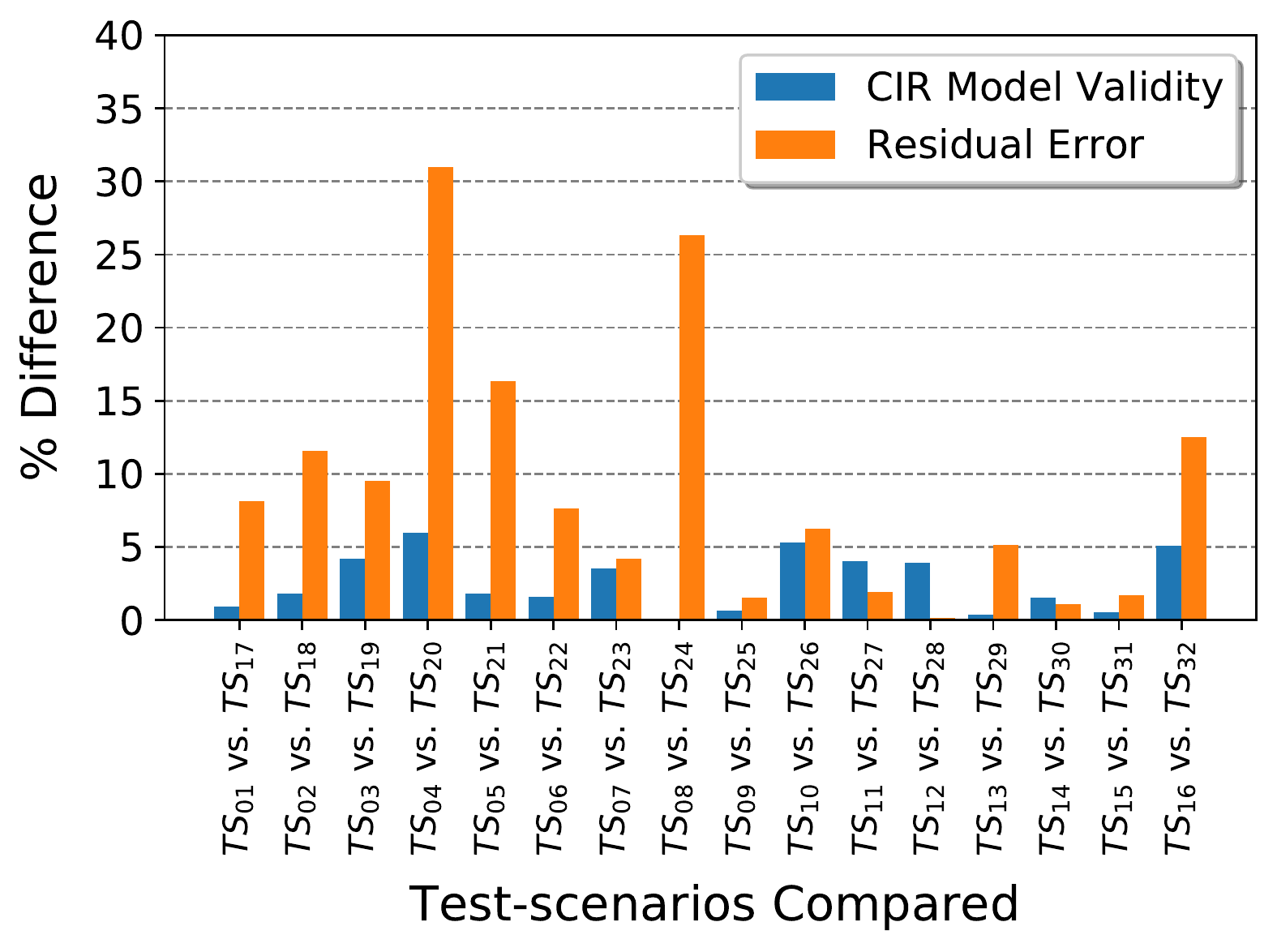}}\hspace*{0.1cm}\hfill%
	\subfloat[802.11ac vs. 802.11n] {\includegraphics[width=.45\linewidth]{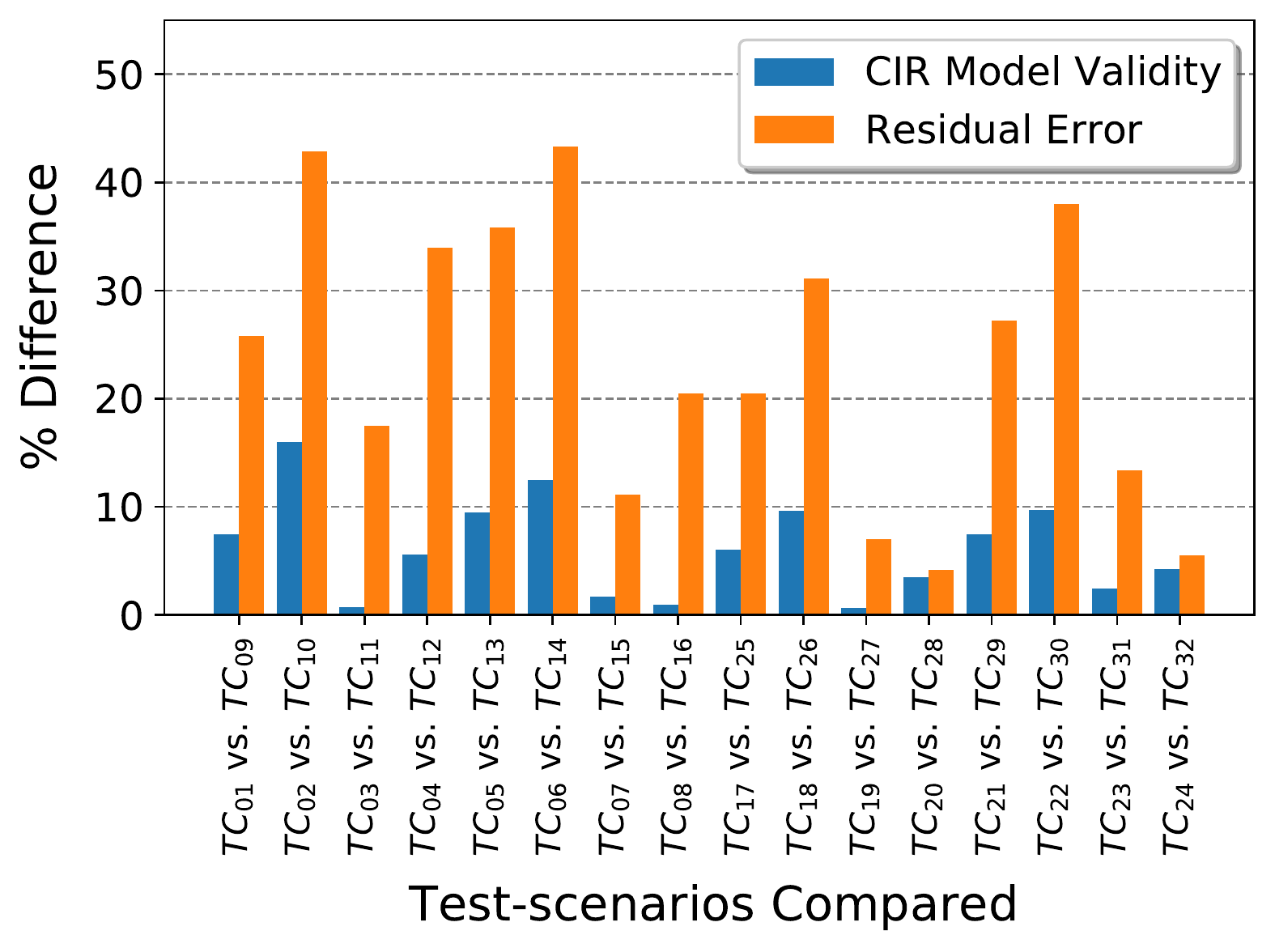}}\hspace*{0.1cm}\hfill%
	\\
	\subfloat[SINR vs. Capacity (P$_{var}$)] 
	{\includegraphics[width=.45\linewidth]{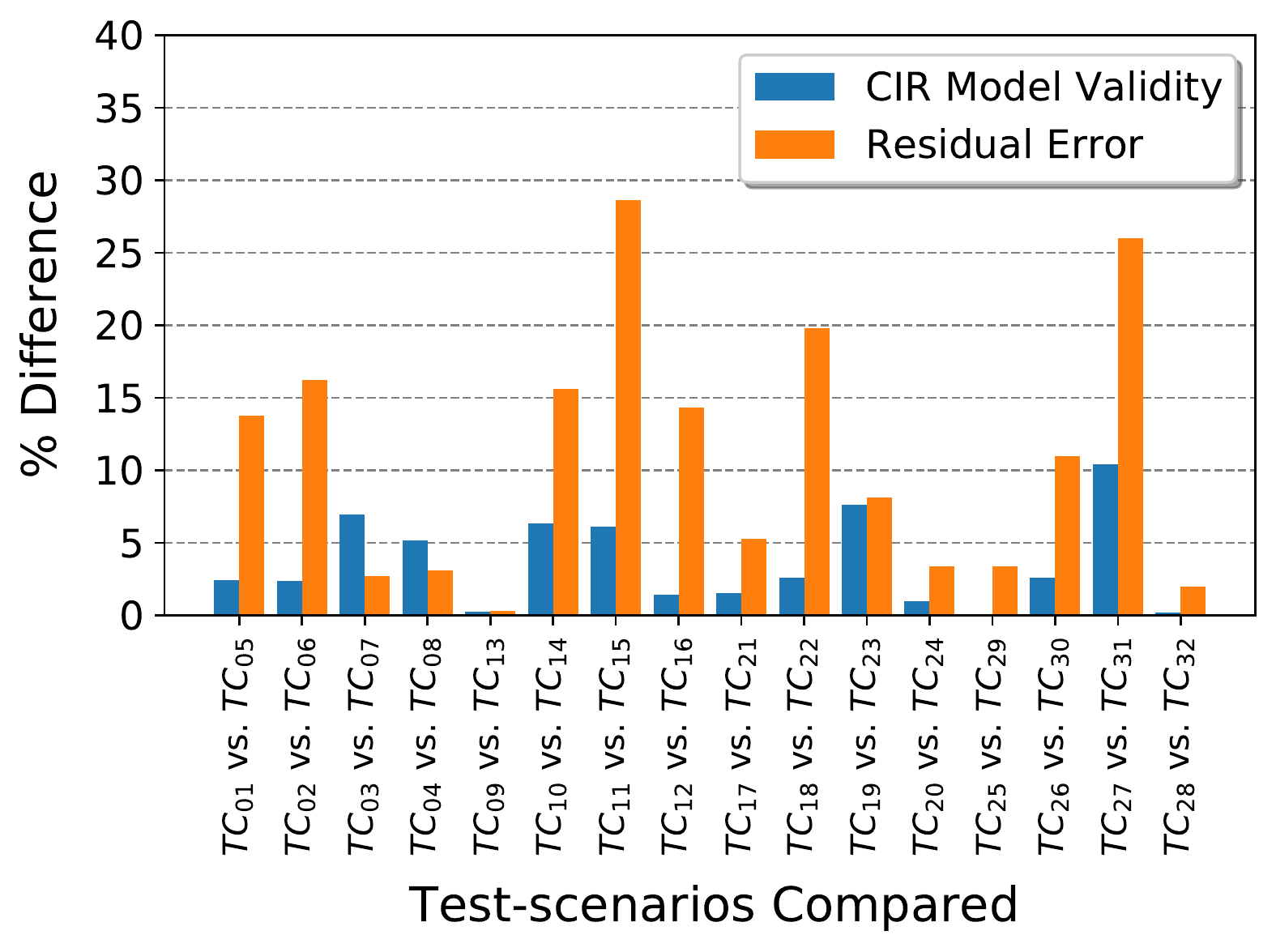}}\hspace*{0.1cm}\hfill
\end{tabular}
  \caption{Test-scenario Specific Comparative Analysis} 
    \label{scenario}
\end{figure*}

\begin{figure*}[ht!]
 \centering%
\begin{tabular}{cc}
    \subfloat[Avg. RMV and Residual Error] {\includegraphics[width=.5\linewidth]{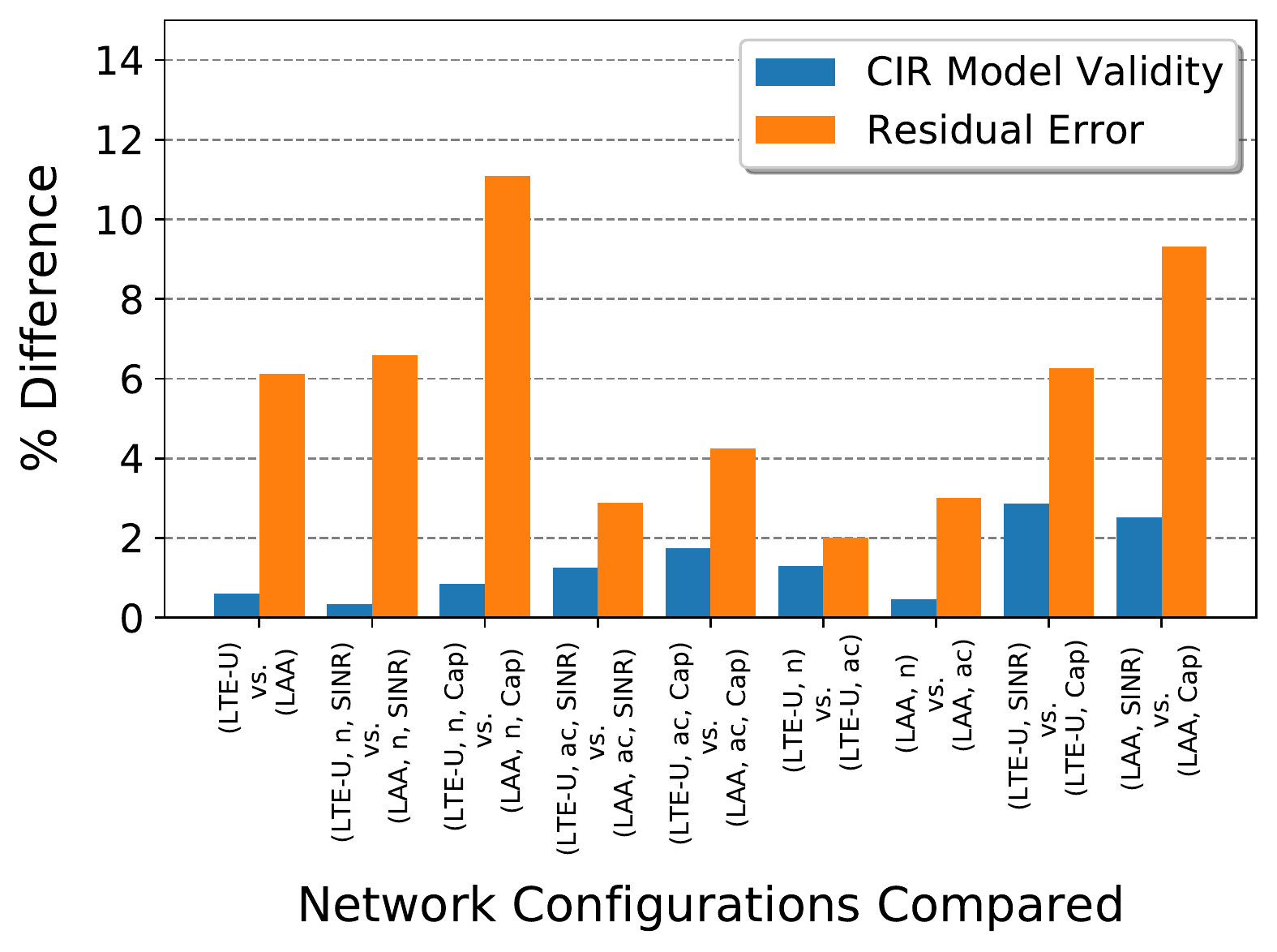}}\hspace*{0.2cm}\hfill%
	\subfloat[Avg. \% Gain \& Std. Deviation in RMV] {\includegraphics[width=.5\linewidth]{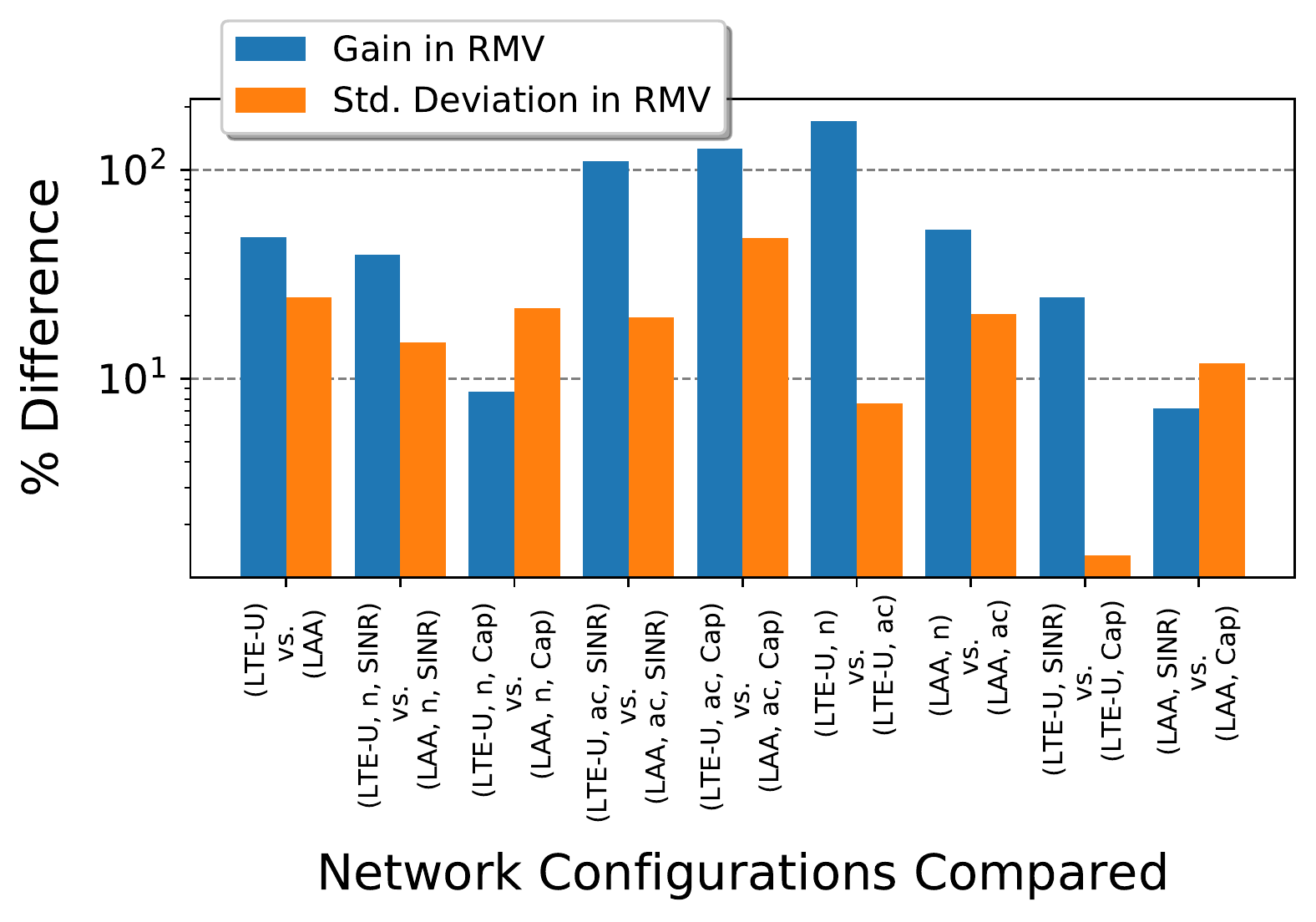}}\hspace*{0.1cm}\hfill%
\end{tabular}
  \caption{Configuration-level Comparative Analysis} 
    \label{average}
\end{figure*}

CIR model parameters are analyzed, and the results are presented for scenario-specific comparisons in Figure~\ref{scenario}, and configuration-level trends in Figure~\ref{average}. Please note that only for Figure~\ref{average}~(b), a logarithmic scale is used to show ``\% Difference" due to a high variation in values. Based on these results, various aspects of unlicensed coexistence network performance are discussed ahead. Some results, such as those related to outliers, are mentioned during the course of the discussion itself.

\subsection{Unlicensed LTE: LTE-U vs LAA}

We begin with measurement based observations on average network capacity, as most comparative studies primarily focus on this metric \cite{UvsLaa}. In 75\% of the test-scenarios, LTE-LAA outperforms LTE-U in coexistence with corresponding Wi-Fi variant (n/ac). Likewise, in 87.5\% scenarios, 802.11ac outperforms 802.11n in coexistence with corresponding LTE variant (LTE-U/LAA). Further, LTE-LAA in coexistence with 802.11n/ac offers a higher SINR on average than LTE-U in all scenarios save one.

The LBT mechanism of LAA is quite similar to the CSMA channel access protocol of Wi-Fi and leads to a higher network capacity on average in LTE-LAA. Further, 
LAA nodes sense the energy level on the medium (-72 dBm) prior to transmission which mitigates co-channel interference from Wi-Fi and other LAA APs, ensuring higher SINR on average than LTE-U. On the contrary LTE-U has a duty-cycle based channel access mechanism which leads to inefficient transmissions and packet-collisions in both, the LTE-U and Wi-Fi components of the coexistence system. 

\par\textbf{Regression Model Validity (RMV)}
LAA and LTE-U models fare equally well, in a scenario specific comparison with $\leq$5\% difference in RMVs in 13/16 comparisons (26/32 scenarios). CIR in LAA seems to be only slightly better as it outperforms LTE-U in the remaining 3 scenarios. In terms of average RMVs across all 32 scenarios, LAA and LTE-U are comparable, although LAA has a slight edge ($<$1\%). Likewise, in LAA-WiFi-Predictor configuration combinations, LAA has a slight edge (0--2\%). 
Prima facie, based on RMV alone, CIR does not seem to be impacted by the unlicensed LTE variant. However, RMV can not be considered to be the sole goodness-of-fit measure for feature relationships. Higher RMV is an indicator of the variation in dependent variable explained by the model, but it does not indicate how far the data-points lie from the regression line. Further, the standard deviation of RMV with kCV for a specific scenario must also be low. The analysis ahead explores these dimensions.

\par\textbf{Residual Standard Deviation (RSD)}
 The capability of a feature relationship model to make accurate predictions is highly desirable for the model to be deployed in real-world network performance management. Thus, residual error or RSD is a measure of precision of the model's predictions and should ideally be low for a robust CIR.

Higher residual error is observed in twice as many LTE-U scenarios as compared to LAA scenarios (5\% margin of error). On average, LTE-U scenarios have a 6\% higher RSD than LAA. Further, average residual error in all LTE-WiFi-Predictor network-configurations is lower for LAA when compared to LTE-U. Thus, LAA models seem to be more precise in their ability to predict coexistence network performance, regardless of the response variable (Capacity or SINR). 
\par\textbf{Gain and Standard Deviation in RMV}
It is important to notice the standard deviation (SD) in CIR model validities when subjected to kCV, especially after LOF outlier removal. While outlier reduction yields higher RMVs, the Gain in RMV should be accompanied with low SD in RMV, averaged across all kCV runs. Thus, we consider high Gain and low SD as a characteristic for stable CIR models.

LTE-U fares much worse than LAA in terms of both Gain and SD. LAA outperforms LTE-U by 47.67\% in Gain and registers a 24.5\% lower SD, averaged across all scenarios. A similar trend can be observed in LTE-WiFi-Predictor combinations as well. Thus, LAA has a higher Gain post-outlier-removal along with a lower SD, which demonstrates robustness of the LAA CIR models.

\par\textbf{Outliers}
For a network system, the outlier \% may be considered to be a good indicator of the degree of fluctuation in network performance, and consequently the ability of a network to deliver the promised Quality of Service (QoS). 
However, the selection of outlier detection algorithm is a subjective choice. While this work steers clear of making inferences based on outliers, we compare the outliers in LTE-U and LAA data detected by LOF algorithm with the outliers detected by ``Minitab," a standard tool for data-analysis \cite{minitab}. Minitab's outlier detection algorithm labels samples with extreme ``leverage points" and ``large residuals" as outliers. As expected the percentage of data-points labeled as outliers is different in LOF and Minitab. However, LTE-U has higher a fraction of outliers as compared to LAA in both LOF (by 9.11\%) and Minitab (by 5.14\%). 

The reason for high fluctuation in LTE-U can be attributed to greater susceptibility of an LTE-U node to the unpredictable interference from Wi-Fi APs in its proximity. This primarily happens during the LTE-U ON state as there are no energy detection thresholds in LTE-U. Unlike LTE-U, Wi-Fi considers the energy threshold as -62 dBm and preamble detection threshold as -82 dBm. Similar to Wi-Fi, the LBT mechanism in LAA has an energy threshold of -72 dBm, making it less vulnerable to interference from Wi-Fi APs, and ensuring fewer extreme network performance fluctuations. Thus, LAA seems to offer a more reliable performance from the perspective of end-user QoS experience. 

\par\textbf{LTE-LAA vs LTE-U: A Feature Relationship Perspective}
A clear pattern emerges after the analysis of various learning model parameters. Residual error, standard deviation in RMV, and outlier \% in LTE-U is higher than LAA, while post-outlier-removal Gain in RMV is lower.
This is true for the majority of test-scenarios regardless of the choice of Wi-Fi variant, predictor variable, and bandwidth allocated. Thus, CIR in LTE-LAA networks is qualitatively better in terms of the spread of data along the expected curve fit. This implies that LAA offers greater consistency in networks performance and lower fluctuations in system variables such as the signal strength or the throughput at the end-user device. 
This finding has a strong correlation with the industry trends. The Global Mobile Suppliers Association (GMSA) report states that 38 operators in 21 countries have made investments in LAA as compared to only 11 operators investing in LTE-U. In terms of global deployments, 30 operators are planning to deploy or are actively deploying LAA networks in 18 countries, in contrast to LTE-U which is being deployed in only 3 countries. Further, LTE-U deployments are designed with an upgrade path to LAA and eLAA \cite{LAA2}.
Clearly, LAA is the preferred choice of industry for LTE unlicensed networks. From a data-learning perspective, this appears to be reasonable as LAA offers a more robust network performance than LTE-U.

\subsection{Wi-Fi: 802.11n vs 802.11ac}
\par\textbf{Measurement Based Analysis} 802.11ac outperforms 802.11n in 87.5\% scenarios in terms of average network capacity. This is expected as 802.11ac supports 80 MHz channels (with optional support up to 160 MHz), higher modulation schemes (256 QAM), and 8x8 Multi-user Multiple-input Multiple-output (MU-MIMO), among other features.

\par\textbf{Feature Relationship Analysis}
802.11ac is slightly better than 802.11n in scenario-specific RMV comparison, while in terms of component-specific average RMV, the two are comparable. The post-outlier-removal Gain in 802.11n is much higher, even though the average RMVs are comparable. However, 802.11ac has a lower deviation in model validities, which implies more reliable CIR models than 802.11n. In terms of residual error, 802.11ac registers lower error in 33\% more models as compared to 802.11n. This signifies more accurate predictive modeling in 802.11ac. 

\par\textbf{802.11ac vs 802.11n : A Feature Relationship Perspective}
The CIR analysis reveals only a marginal advantage in coexistence performance for 802.11ac as compared to 802.11n. The trends are underwhelming because the 802.11ac standard supports compressed \textit{beamforming} which along with channel state information (CSI) is quite efficient in mitigating link-conflicts \cite{ac}. Hence, a stronger relationship between network capacity and SINR was expected.

However, the observations can be reasonably explained through two facts. First, in an LTE-WiFi coexistence system, the unlicensed LTE (LTE-U/LAA) subsystem has a greater impact on the performance of the Wi-Fi subsystem than the latter has on the former. Thus, the unlicensed LTE subsystem is the primary determinant of the overall system performance. Second, the adverse impact of LTE-U on coexisting Wi-Fi (n/ac) performance is much worse than that of LAA on Wi-Fi \cite{UvsLaa}. The duty cycling mechanism of LTE-U combined with the LTE-U's transmission at energy threshold's lower than those prescribed by Wi-Fi cause interference to Wi-Fi transmissions \cite{LTEU-WiFi1}. LAA's LBT avoids collisions with Wi-Fi transmissions, and leads to a better coexistence system performance. This is observed in the LTE-WiFi-Predictor combination analysis as well.

Thus, from a data analysis perspective, the unlicensed LTE is the dominant subsystem in the coexistence paradigm, and determines the overall system performance. Further, the feature relationship analysis of network-data also supports the findings from measurement based studies that LTE-U has a higher adverse impact on Wi-Fi performance as compared to LAA \cite{UvsLaa}. Another major takeaway is that it seems more appropriate to study the Wi-Fi (n/ac/ax) subsystems performance only in conjugation with the coexisting unlicensed LTE (LTE-U/LAA) or 5G NR-U subsystem.

\subsection{Choice of Network Predictor Variable}
A bidirectional regression analysis reveals the impact that the choice of predictor variable \emph{e.g.,} SINR (P$_{SINR}$) or Capacity (P$_{Cap}$), has on network feature relationships. We find that network capacity is a much better predictor of SINR than SINR is of network capacity. This is a pattern that can be clearly and consistently seen across all CIR model parameters and all comparative themes without any ambiguity. In scenario-specific comparison, RMV of P$_{SINR}$ models is always either comparable to, or lower than P$_{Cap}$ models. RMV of P$_{Cap}$ models is higher on average for both LTE-U and LAA components when compared to RMV of corresponding P$_{SINR}$ models. P$_{Cap}$ models also exhibit a significantly higher post-outlier-removal Gain and lower average standard deviation in RMV. Finally, the residual error is higher in P$_{SINR}$ on average, and in twice as many scenarios, when compared to P$_{Cap}$.

It may seem counter-intuitive that it is more accurate to predict the expected values of SINR for given values of network capacity, than the reverse. However, recent analysis of operator data gathered from public LAA deployments shows that high SINR doesn't always guarantee high throughput in coexistence deployments, as end-user QoS depends on other factors such as resource block allocation \cite{icdcn}. On the other hand, for high throughput a high SINR is a necessary, if not a sufficient condition.

Thus, the direction of NFR analysis and the choice of predictor has a clear effect on the learned network model, regardless of the unlicensed LTE and Wi-Fi variants considered. Further, this also indicates that other variables may also be relevant to the unlicensed coexistence NFR analysis such as resource block allocation, physical cell-id, \emph{etc.}
\subsection{Impact of Bandwidth}
\begin{figure*}[ht!]
 \centering%
\begin{tabular}{cc}
    \subfloat[RSD and Bandwidth] {\includegraphics[width=.5\linewidth]{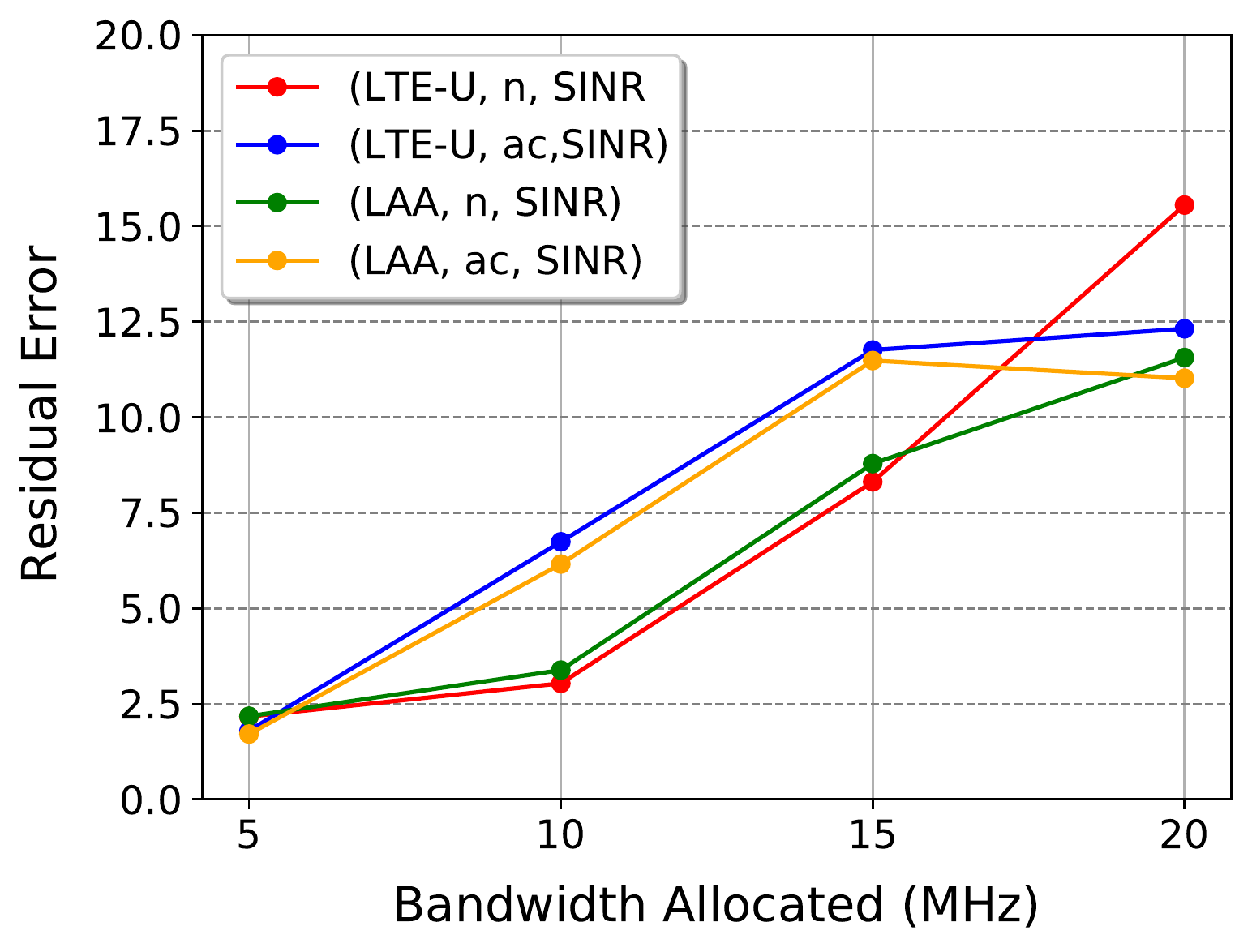}}\hspace*{0.2cm}\hfill%
	\subfloat[Normalized RSD and Bandwidth] {\includegraphics[width=.5\linewidth]{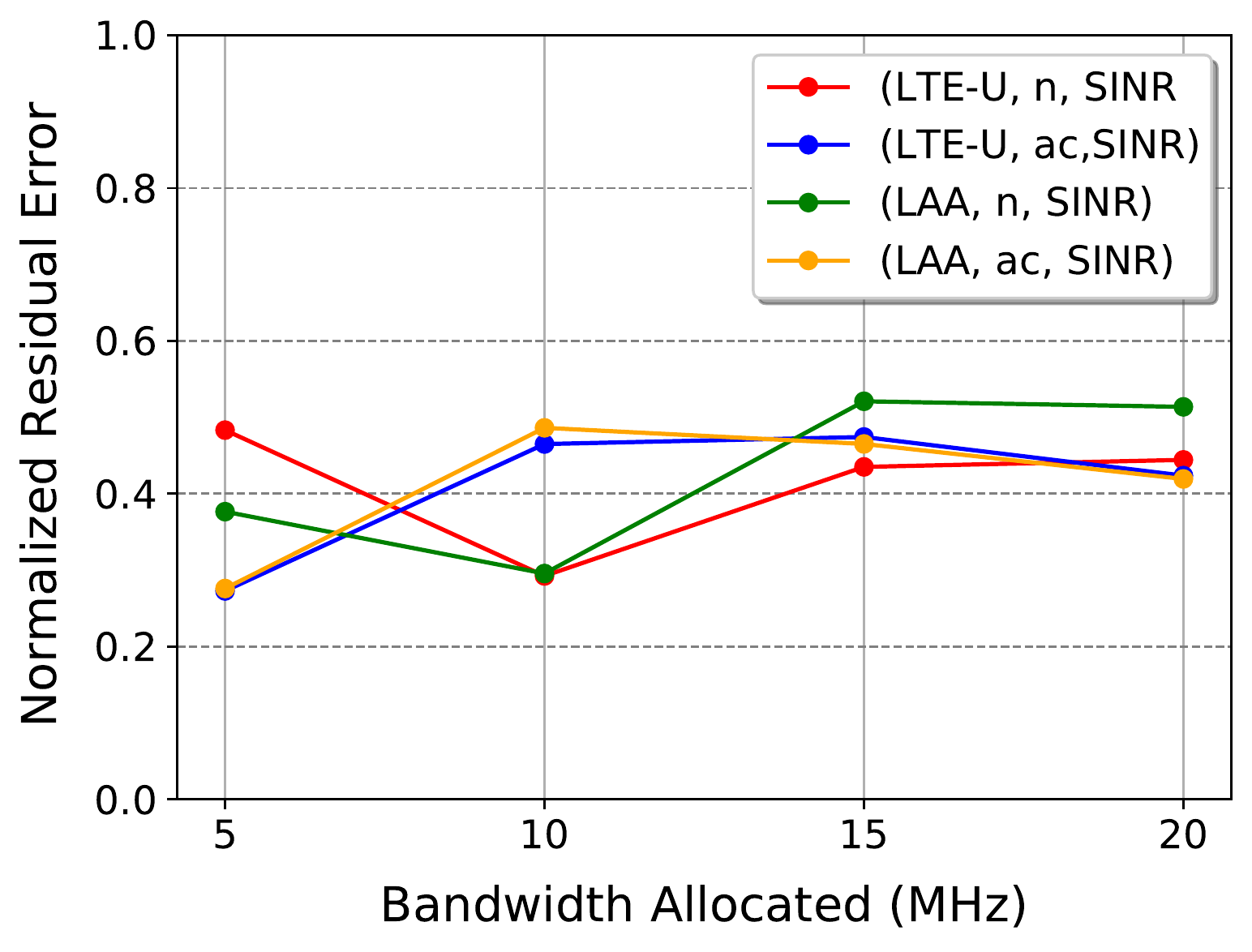}}\hspace*{0.1cm}\hfill%
\end{tabular}
  \caption{Impact of Bandwidth} 
    \label{bandwidth}
\end{figure*}

From Figure~\ref{bandwidth}~(a), prima facie it appears that when throughput is the response variable, the residual error of the models increases consistently with bandwidth. This pattern seems consistent for both LTE-U and LAA models. This would make sense as well, because with higher bandwidth allocation there is a greater possibility of fluctuation in network capacity values in real-world systems due to poor resource allocation and temporal variation in factors such as interference.

To confirm this pattern, we normalized the coexistence data and learned the feature relationships and associated parameters again. The data was normalized as $\hat{\mathbf{z}} = \frac{\mathbf{z} - \mathbf{\mu}}{\mathbf{\sigma}}$, where $\mathbf{\mu}, \sigma$ are the mean and the standard deviation of the data. As a result, the processed data is zero mean and unit variance, and thus more suited to evaluate the impact of bandwidth.
 Prior to normalization, in 11 out of 12 scenario-specific comparisons the RSD had increased with an increase in bandwidth. However, after normalization, in almost half the scenarios there is no increase in residual error with increase in bandwidth and the earlier trend is non-existent.
 
 This finding has serious implications for QoS promised to the end-user. Cellular operators attempt to satisfy the guaranteed user demand according to the data plan. Had higher bandwidth allocation exhibited an association (if not causation) with greater fluctuation in network performance, it would be worrisome. However, this does not seem to be the case.
 
\begin{figure}[t!]
                \centering
                \includegraphics[width=0.95\linewidth,trim=5mm -5mm 0mm -3mm]{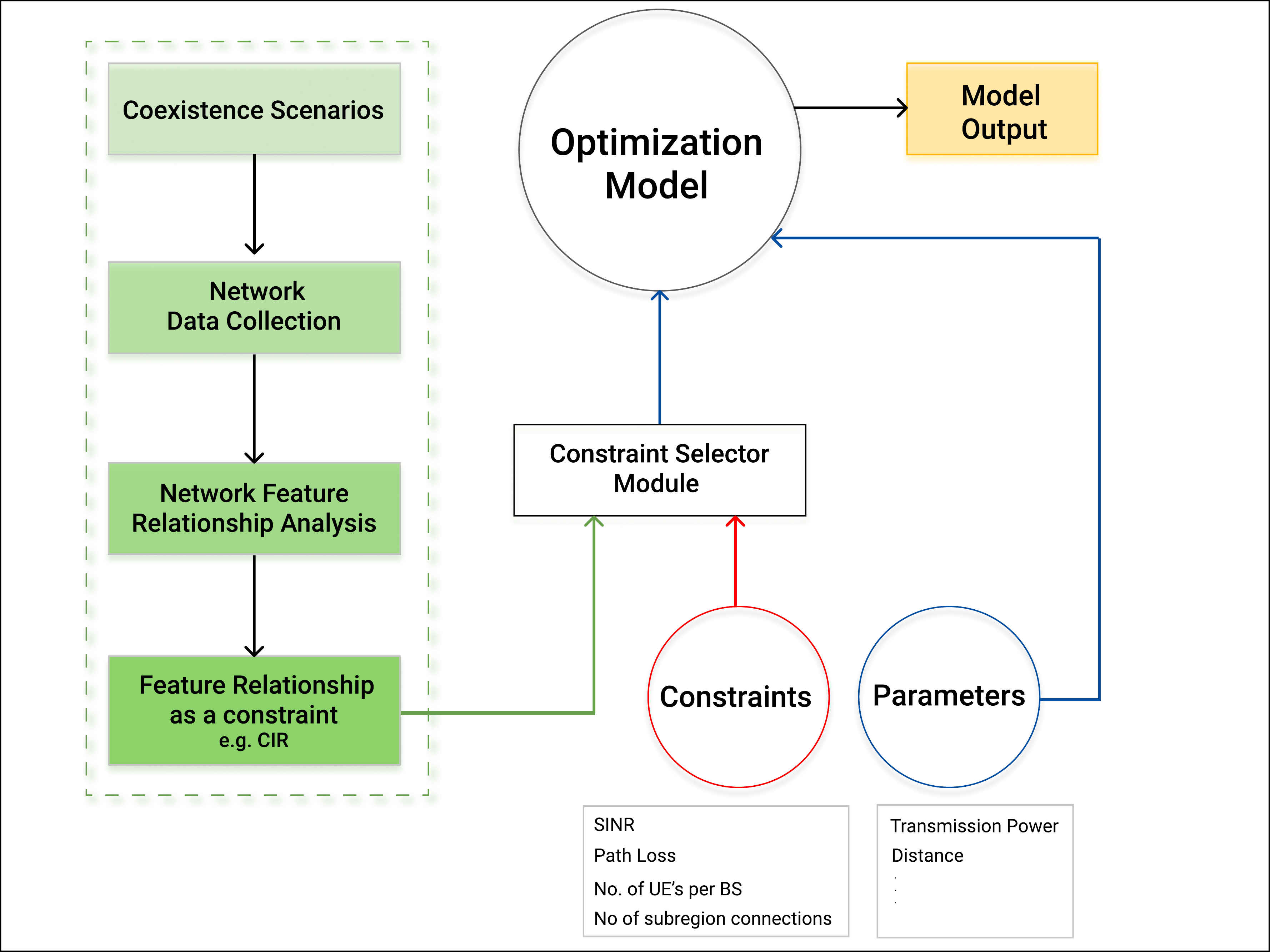}
                \caption{Network Feature Relationship based Optimization}
                \label{nefro}
        \end{figure}

\section{NFRs and Dense Network Optimization}
 The feature relationships learned from network-data can be further utilized in improving dense network performance. 
 
\subsection{Network Feature  Relationship  based  Optimization (NeFRO)}
To facilitate the use of NFRs in network performance enhancement, this work proposes the \textit{\textbf{Network Feature Relationship based  Optimization}} (NeFRO) framework. The high-level schema of the NeFRO framework is outlined in Figure~\ref{nefro}. 
First, data is collected for a network deployment periodically. In each epoch, network feature relationship analysis is performed using machine learning algorithms. Strong NFRs are identified and selected for possible utilization in network performance optimization. These NFRs are fed to a \textit{constraint selector module} that selects relevant constraints for the optimization model/formulation. The module compares an NFR learned from network data for a network feature-point set $\{f_1, f_2,\ldots,f_n\}$ with available theoretical constraints relevant to the feature point set. While the NFR is more ``suitable," as it is derived from actual network data, it still has to be tested for \textit{convergence time viability}. Thus the constraint selector module compares the NFR with the theoretical constraint for complexity, and selects the more viable constraint for network optimization. 
Although the illustration highlights the process-flow for a coexistence network, the NeFRO approach will apply similarly to network optimization in all wireless networks, with minor modifications, if required. 
\par\textbf{Benefits of the NeFRO Approach}
There are several advantages of the proposed NeFRO framework over conventional network optimization. First, since the learned NFRs are grounded in empirical data, they reflect the ambient network conditions. Therefore, it is more practical to use them in network performance optimization than theoretical constraints involving similar network variables. Second, NFRs can be used ``as is" in optimization without making any assumptions, unlike theoretical constraints which need to be justified through assumptions. Finally, if the learned NFRs are less complex than the theoretical constraints, it automatically solves the problem of arbitrary or forced relaxation of constraints. Even if the NFRs are of a comparable complexity and require similar computational overhead, they have the advantage of reflecting the actual network parameter dynamics, which facilitates a more informed network optimization. 

\subsection{Implementation and Validation of NeFRO}
Convergence time and accuracy trade-off is a primary challenge in dense network performance optimization \cite{dense5}. Therefore, NeFRO envisions the twin objectives of \textit{convergence time reduction}, while maintaining high \textit{accuracy}, vis-à-vis the baseline optimization model. The validation of NeFRO is done by implementing it on recent state-of-the-art studies on coexistence network optimization.
\par\textbf{Validation Methodology}
The validation methodology involves the following steps. First, works with two optimization objectives are considered, \emph{viz.} network signal strength optimization and network capacity optimization. The proposed optimization models are implemented on GAMS \cite{GAMS}, as per the network configuration and specifications of the testbed/experiments. Second, the  baseline optimization models are implemented for the test-scenarios considered in this work. Further, two values are observed, (a) the optimal value of network performance metric (SINR or Capacity), and (b) the \textit{convergence time} required by the formulation to arrive at the optimal value. Thereafter, the complex theoretical SINR-Capacity constraint in each of the proposed optimization formulation is replaced with the second-degree polynomial CIR equation derived from feature relationship analysis in this work. Please note that the baseline models that optimize network capacity are considered for test-scenarios with SINR as the predictor, and vice-versa.
\par\textbf{Evaluation of NeFRO}
Two yardsticks are considered to carry out the performance evaluation of NeFRO. First, is the closeness of the ``NeFRO Optimal" value generated by the NeFRO model, to the optimal value generated by the baseline literature model. This is referred to as the \textbf{Accuracy} of the NeFRO model. Accuracy can be defined as, the \textit{```\% difference in the optimal value generated by the baseline model and the NeFRO-optimal value."} Second, is the reduction in the time taken by the NeFRO model to arrive at the optimal value. This is defined as \textbf{Convergence Time Fraction} (CTF). CTF indicates \textit{``what fraction (\%) of the baseline model's convergence time is NeFRO's convergence time."} \footnote{For example, if baseline model takes $10ms$ to converge at the optimal solution, and NeFRO requires $9ms$ to arrive at the NeFRO-optimal value, then CTF is 90\%}

Thus, NeFRO is evaluated on its ability to offer a \textit{low CTF} while maintaining \textit{high Accuracy}, with respect to the baseline optimization model. Please note that the state-of-the-art optimization models are implemented for the small-scale dense unlicensed coexistence scenarios implemented on the experimental testbed. We expect that in a real-world network of a much higher scale and density, the benefits of NeFRO will be far more pronounced.

\par\textbf{Baseline Optimization Models Considered}
Four recent works are considered that propose formulations to optimize coexistence network performance. Two of these works aim at optimizing network capacity, while the other two optimize signal strength available to the UEs. 
A brief description is presented, starting with the capacity optimization works. An optimal resource allocation scheme aimed at maximizing LTE-LAA capacity in a LTE-WiFi coexistence network is proposed in \cite{Cap1}. Another study proposes an LBT-compliant channel access approach for both LTE-U/LAA in the 5GHz band that seeks to maximize system throughput, while also mitigating the impact of interference from the unlicensed LTE on the Wi-Fi subsystems capacity \cite{Cap2}. Further, \cite{SINR2} seeks to enhance and optimize network signal strength for LTE-U/LAA coexistence networks through strategic optimal placement of nodes. Finally, the model proposed in \cite{SINR1}, aims to optimize network performance by taking into account the spectrum usage of Wi-Fi APs in addition to the optimal placement of nodes. Henceforth, the capacity optimization models \emph{viz.,} \cite{Cap1} and \cite{Cap2}, are referred to as COM$_1$ and COM$_2$, respectively. Likewise, signal-strength optimization models \emph{viz.,} \cite{SINR1} and \cite{SINR2}, are referred to as SOM$_1$ and SOM$_2$, respectively. 
\begin{figure*}[ht!]
 \centering%
\begin{tabular}{cc}
    \subfloat[NeFRO vs. COM$_1$ ] {\includegraphics[width=.45\linewidth]{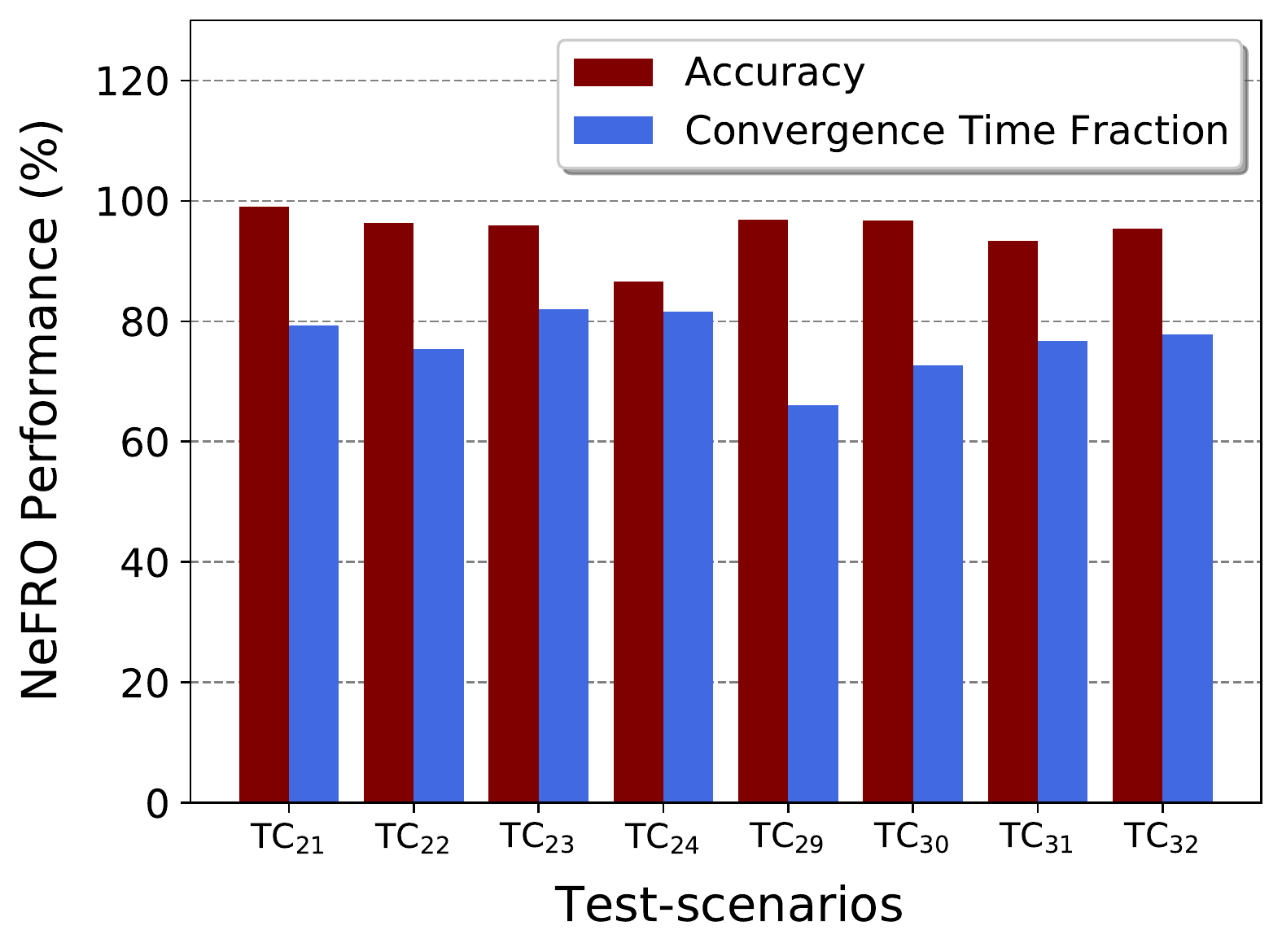}}\hfill%
	\subfloat[NeFRO vs. COM$_2$] {\includegraphics[width=.45\linewidth]{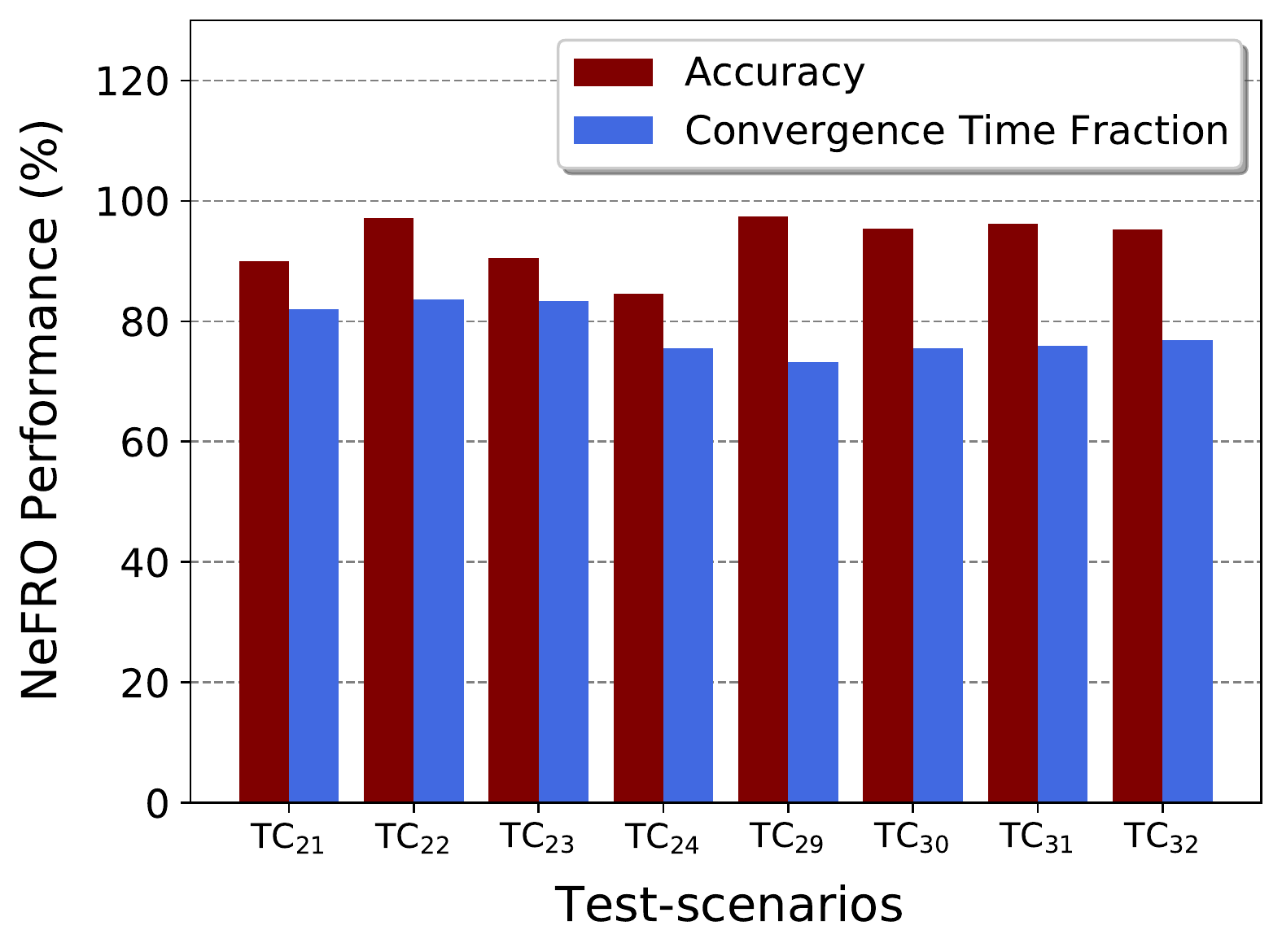}}\hfill
	\\
	  \subfloat[NeFRO vs. SOM$_1$] {\includegraphics[width=.45\linewidth]{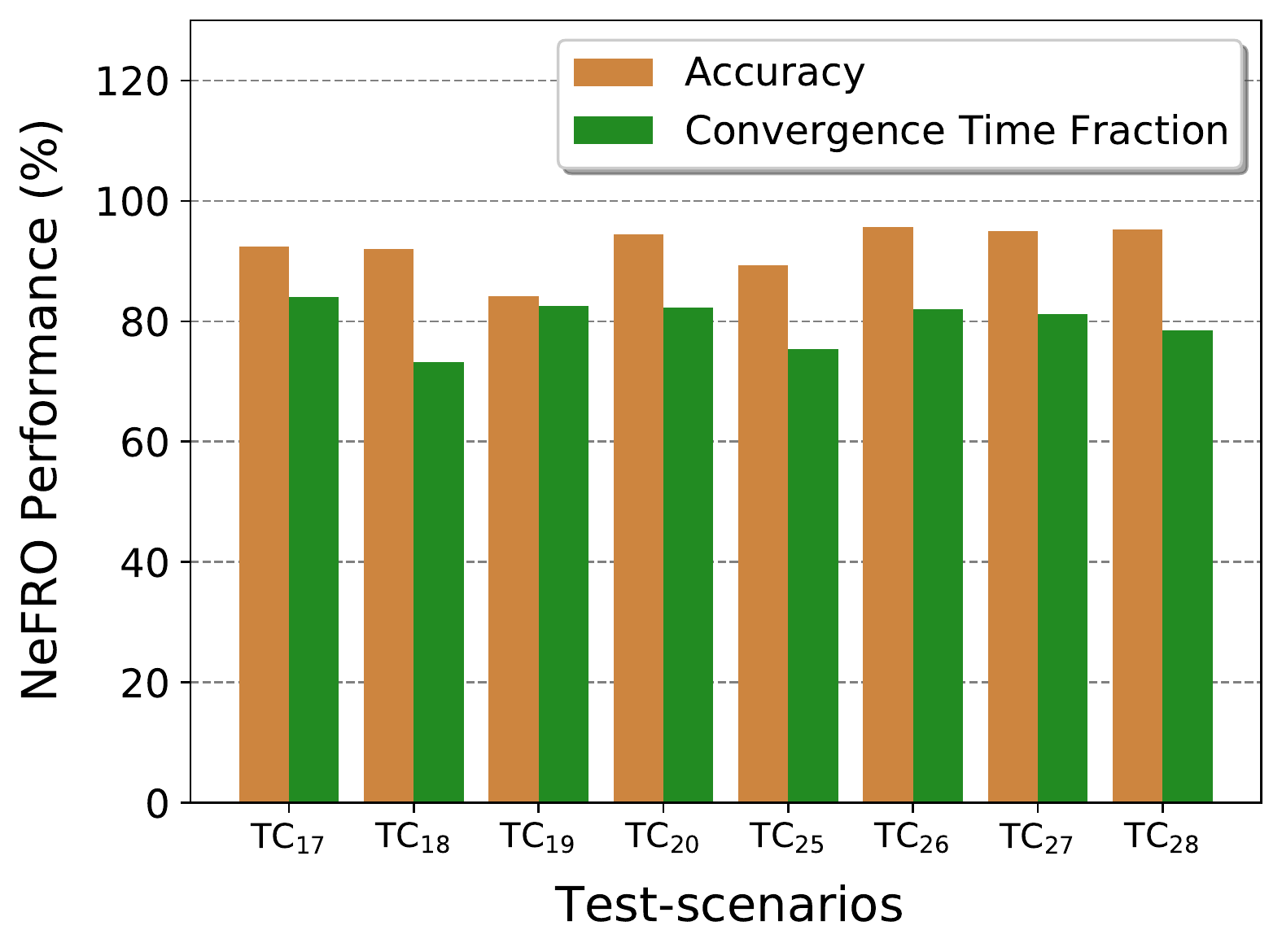}}\hfill%
	\subfloat[NeFRO vs. SOM$_2$] {\includegraphics[width=.45\linewidth]{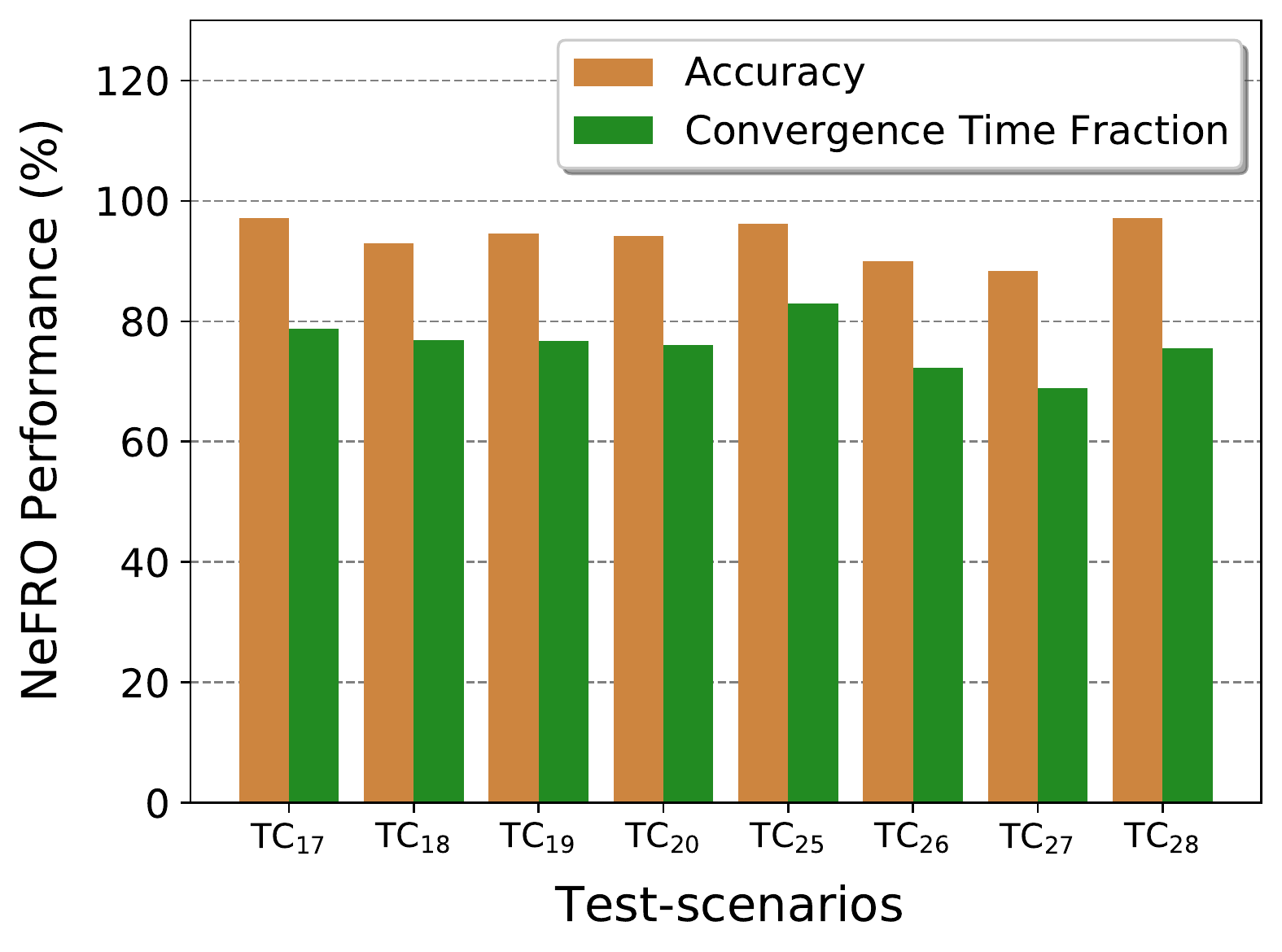}}\hfill
\end{tabular}
  \caption{NeFRO Performance in LAA Capacity and SINR Optimization} 
    \label{nefroLAA}
\end{figure*}

\subsection{Optimization Results and Performance Evaluation}

\begin{figure*}[ht!]
 \centering%
\begin{tabular}{cc}
    \subfloat[NeFRO vs. COM$_1$ ] {\includegraphics[width=.45\linewidth]{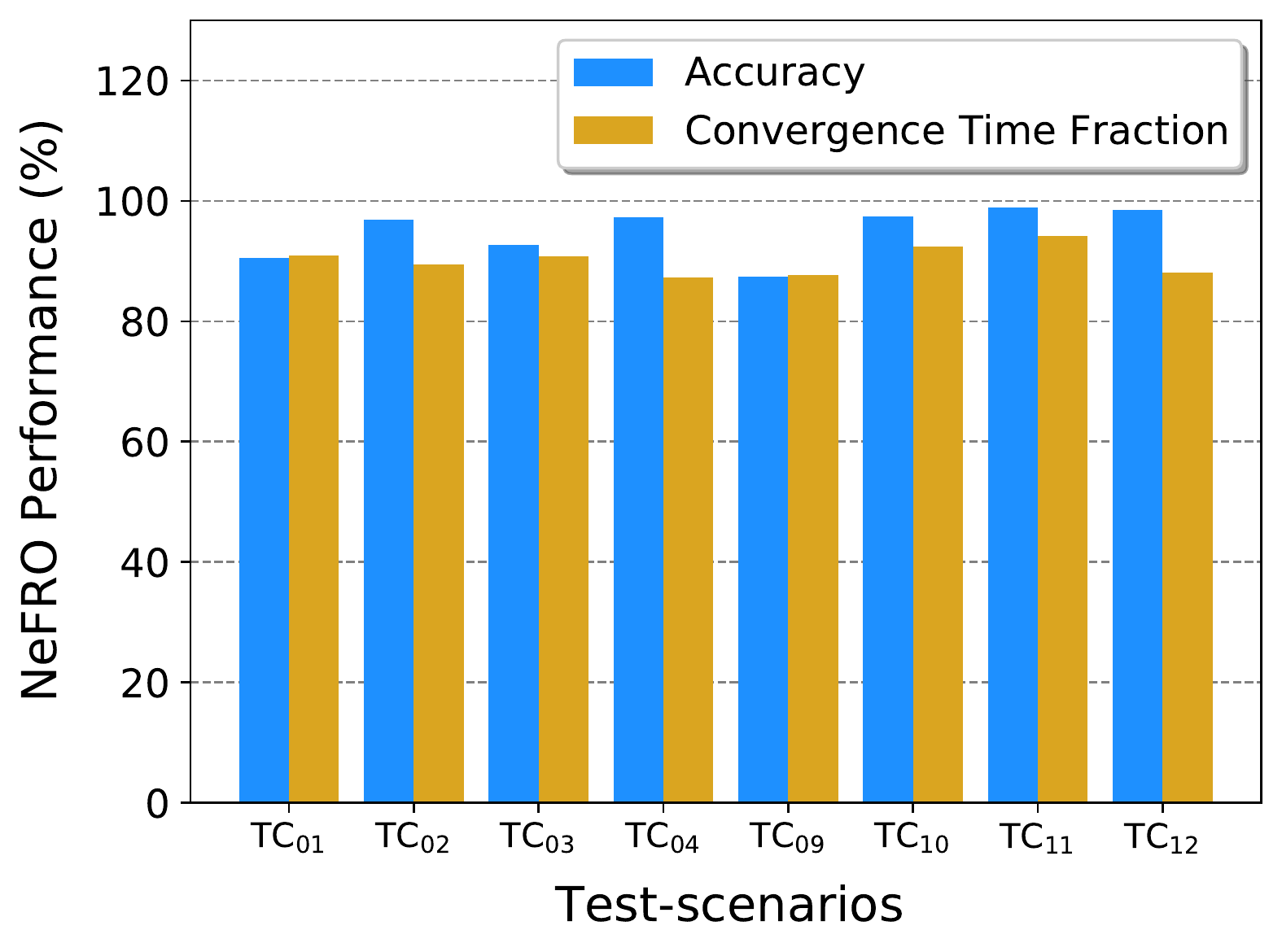}}\hfill%
	\subfloat[NeFRO vs. COM$_2$] {\includegraphics[width=.45\linewidth]{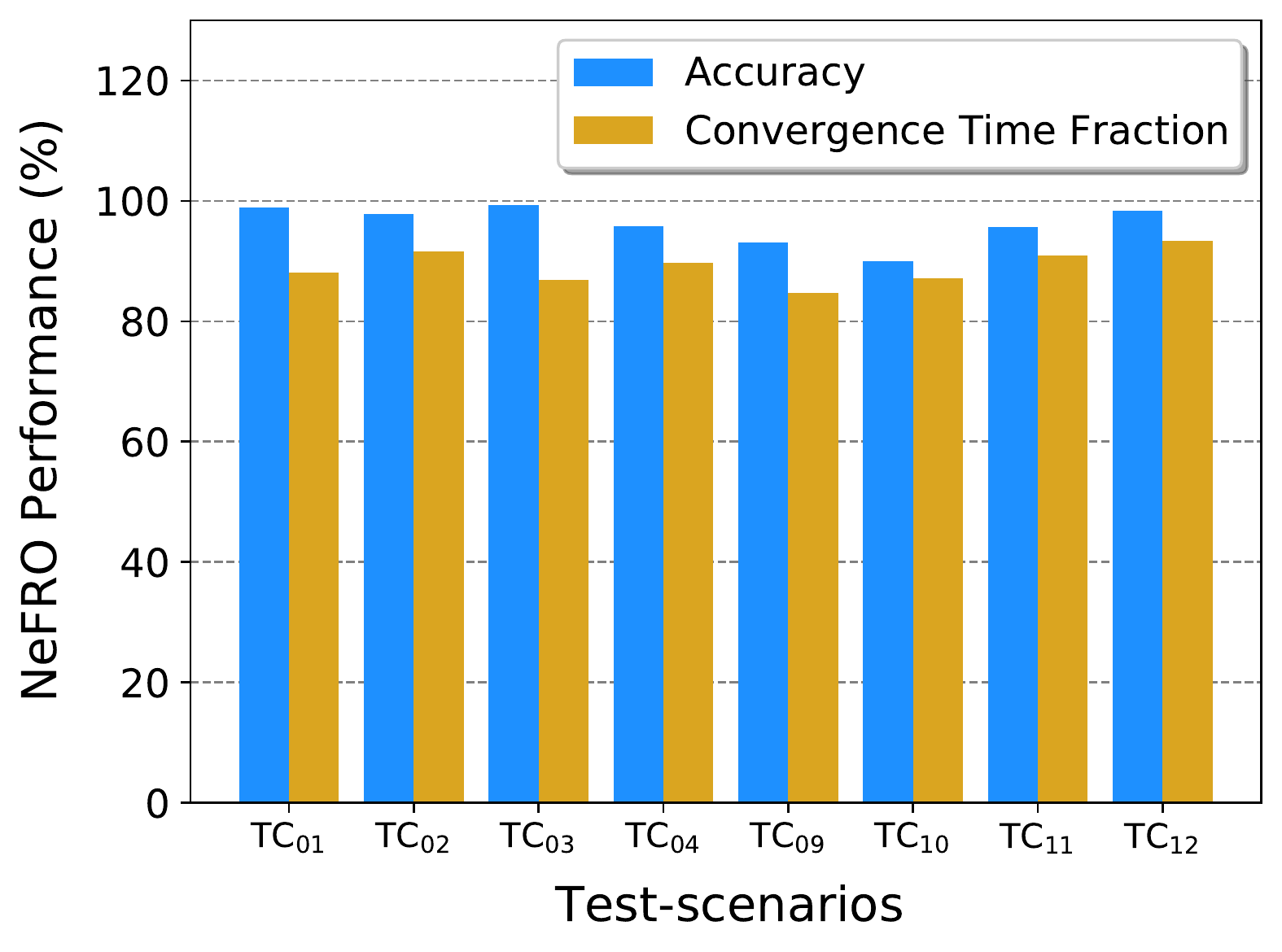}}\hfill
	\\
	  \subfloat[NeFRO vs. SOM$_1$] {\includegraphics[width=.45\linewidth]{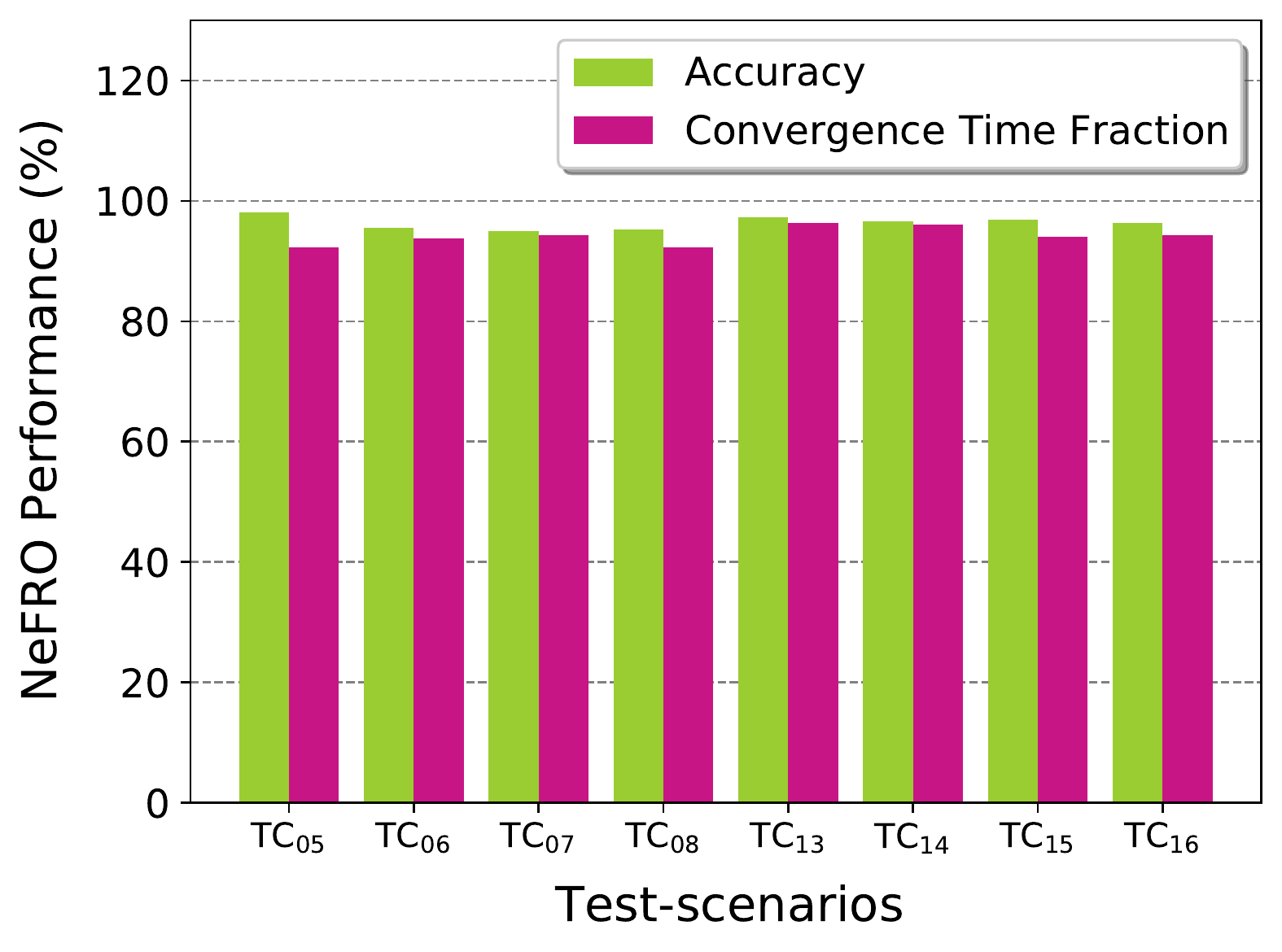}}\hfill%
	\subfloat[NeFRO vs. SOM$_2$] {\includegraphics[width=.45\linewidth]{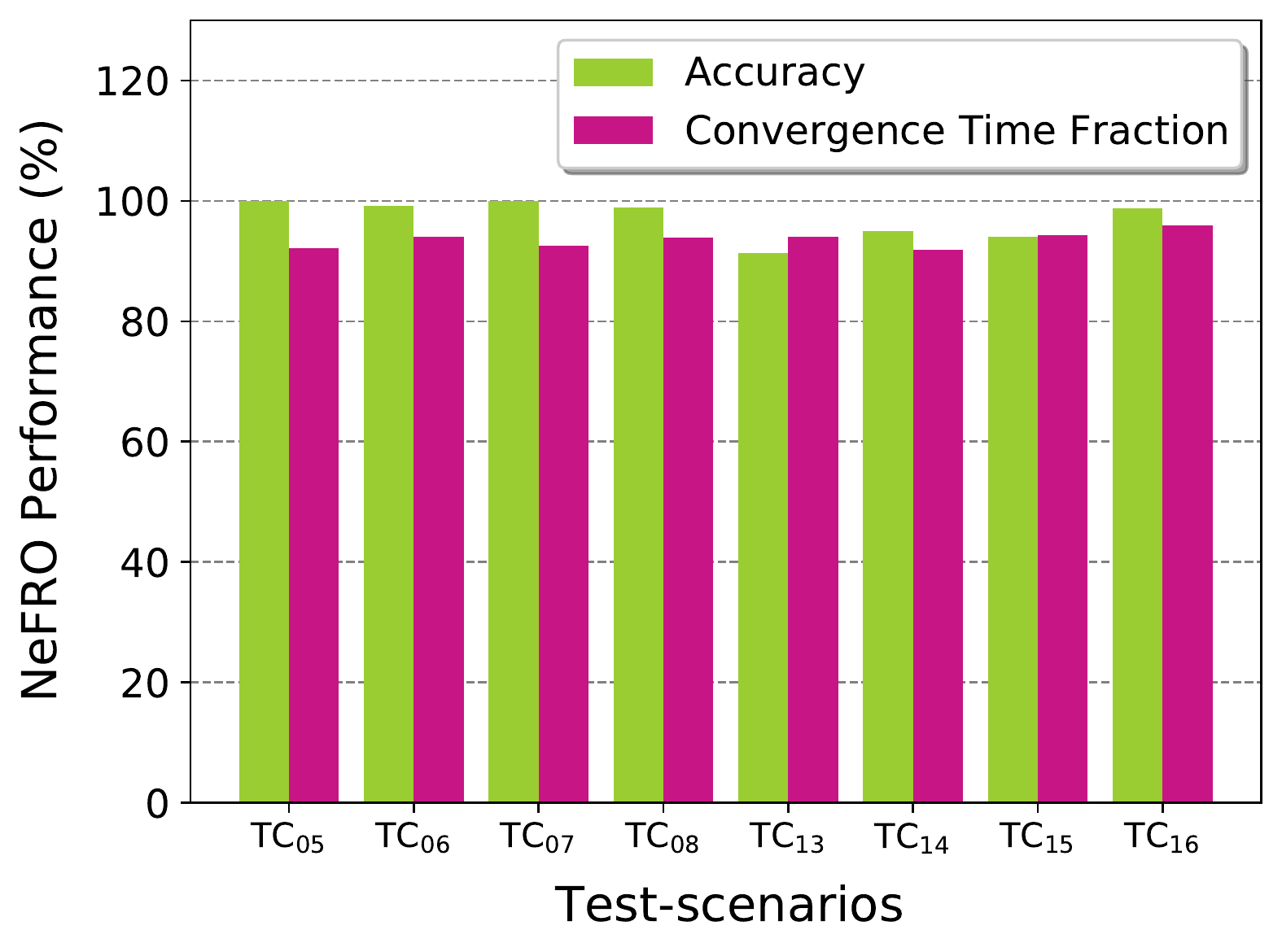}}\hfill
\end{tabular}
  \caption{NeFRO Performance in LTE-U Capacity and SINR Optimization} 
    \label{nefroLTU}
\end{figure*}

The results of the optimization simulations run on GAMS are presented in Figure~\ref{nefroLAA} and Figure~\ref{nefroLTU}, for LAA and LTE-U test-scenarios, respectively. Further, Figures~\ref{nefroLAA}~(a), \ref{nefroLAA}~(b), \ref{nefroLTU}~(a), and \ref{nefroLTU}~(b), present results for test-scenarios where the objective is to optimize network capacity. The remaining figures show results for signal-strength optimization test-scenarios.  

It can be discerned that NeFRO performs remarkably well by reducing the required convergence times while delivering NeFRO-optimal values very close to the optimal results of the respective models. A scenario-specific evaluation of NeFRO can be performed by observing the difference in the length of bars of Accuracy and CTF for a particular test-scenario. The greater the difference in their height, the lower is the trade-off, and the better is the NeFRO performance. 
Two points are noteworthy. First, in LAA scenarios NeFRO offers a significant reduction in convergence time, while in LTE-U scenarios, the CTF is somewhat subdued. Network optimization in LTE-U is inherently more challenging due to its channel access mechanism. Hence, it is more computationally intensive, and requires a longer convergence time. Second, for LAA scenarios the difference in NeFRO performance for capacity optimization and SINR optimization is negligible. However, in LTE-U, there appears to be a difference in NeFRO performance for these two objectives. Particularly, the CTF for SINR optimization in LTE-U is rather low. 

The average performance of NeFRO across all test-scenarios for the four optimization models is presented in Table~\ref{nefroT}. On average, when compared to SOM$_1$ and SOM$_2$, the CTF of NeFRO is lower than its average Accuracy, showing a marginal gain. However, Figure~\ref{nefroLTU}~(d) shows that for two scenarios there seems to be no overall gain from NeFRO as compared to SOM$_2$. Thus one dimension that needs to be further investigated is the variation in accuracy and convergence time-trade off with application of NeFRO. It is possible that the correlation or association between the RMV of the learned model and the network performance metric which is the objective of the optimization (SINR or Capacity), may explain this variation.

In general, NeFRO outperforms the baseline model across all test-scenarios, and both unlicensed LTE variants, by significantly reducing the convergence time. The average Accuracy, as shown in Table~\ref{nefroT}, is very high as well. Further, NeFRO seems to perform better in LTE-LAA scenarios as compared to LTE-U, which can be expected based on the discussion and findings presented in this work. Thus, the NeFRO framework stands validated.

\begin{table} [hbt]
\caption{Performance Trends in Test-scenarios}
\centering
\small
\begin{tabular}{|m{1.5cm}|m{0.85cm}|m{0.85cm}|m{0.85cm}|m{0.85cm}||m{0.85cm}|m{0.85cm}|m{0.85cm}|m{0.85cm}|}
\hline 
\multicolumn{1}{|c|}{\textbf{NeFRO}}&\multicolumn{4}{|c||}{\textbf{LTE-LAA Scenarios} ($\%$)}&\multicolumn{4}{|c|}{\textbf{LTE-U Scenarios} ($\%$)}\\ \cline{2-9}
\multicolumn{1}{|c|}{\textbf{Parameter}}&\textit{COM$_1$}&\textit{COM$_2$}&\textit{SOM$_1$}&\textit{SOM$_2$}&\textit{COM$_1$}&\textit{COM$_2$}&\textit{SOM$_1$}&\textit{SOM$_2$}\\
\hline  
CTF&76.46&78.25&79.89&76.02 &90.10&89.05&94.17&93.60\\
\hline 
Accuracy&95.04&93.31&92.28&93.82 &94.97&96.12&96.38&97.16\\
\hline  
\end{tabular} 
\label{nefroT}
\vspace{-0.1cm}
\end{table}

\section{Conclusion and Future Direction}
This work presented a comparative study of unlicensed coexistence networks through network feature relationship analysis. Network-data was collected through comprehensive real-world experiments and then analyzed through a family of regression algorithms. The relevance of network feature relationships was highlighted by analyzing LTE-WiFi networks on a variety of regression model parameters such a R-sq, residual error, \emph{etc.} Several insightful inferences were made on aspects such as the impact of bandwidth, residual error, and outliers on coexistence network performance. Further, NeFRO, a feature relationship based optimization framework was proposed and validated through signal strength and capacity optimization. NeFRO offered reduced convergence times by as much as 24\% and offered accuracy as high as 97.16\% on average. 

In the future, we will investigate convergence time and accuracy trade-off by considering feature relationships of varying degrees. Further, studying the association between the R-sq of the learned models and the network performance metrics is also a relevant topic. The impact of control/signaling data on network feature relationships will be explored as well. Most importantly, we intend to implement an AR system on a simulator and employ NeFRO to reduce latency.

\bibliographystyle{splncs04} 
\bibliography{ref_journal}

\end{document}